\documentclass[acmlarge,nonacm,screen]{acmart} 
\usepackage{graphicx}
\usepackage{subcaption}
\usepackage{booktabs} 
\usepackage{array}    
\usepackage{soul}
\usepackage{enumitem}
\usepackage[section]{placeins}

\AtBeginDocument{%
  }

\begin{document}

\title{The CTSkills App - Measuring Problem Decomposition Skills of Students in Computational Thinking}

\author{Dorit Assaf}
\authornote{Corresponding authors}
\email{dorit.assaf@fhnw.ch}
\orcid{0000-0002-7877-3041}
\affiliation{%
  \institution{University of Applied Sciences and Arts Northwestern Switzerland (FHNW), School of Education}
  \city{Windisch}
  \country{Switzerland}
}

\author{Giorgia Adorni}
\authornotemark[1]
\affiliation{%
  \institution{Università della Svizzera italiana (USI), Dalle Molle Institute for Artificial Intelligence (IDSIA)}
  \city{Lugano}
  \country{Switzerland}}
\email{giorgia.adorni@usi.ch}
\orcid{0000-0002-2613-4467}

\author{Elia Lutz}
\authornotemark[1]
\orcid{0009-0007-0011-3987}
\email{elia.lutz@fhnw.ch}
\affiliation{%
  \institution{University of Applied Sciences and Arts Northwestern Switzerland (FHNW), School of Education}
  \city{Windisch}
  \country{Switzerland}
}

\author{Lucio Negrini}
\email{lucio.negrini@supsi.ch}
\orcid{0000-0001-6793-6258}
\author{Alberto Piatti}
\email{alberto.piatti@supsi.ch}
\orcid{0000-0002-5196-4630}
\affiliation{%
  \institution{University of Applied Sciences and Arts of Southern Switzerland (SUPSI), Department of Education and Learning (DFA)}
  \city{Locarno}
  \country{Switzerland}}

\author{Francesco Mondada}
\affiliation{%
  \institution{Ecole Polytechnique Fédérale de Lausanne (EPFL), Mobile Robotic Systems Group (MOBOTS)}
  \city{Lausanne}
  \country{Switzerland}}
\email{francesco.mondada@epfl.ch}
\orcid{0000-0001-8641-8704}

\author{Francesca Mangili}
\affiliation{%
  \institution{University of Applied Sciences and Arts of Southern Switzerland (SUPSI), Dalle Molle Institute for Artificial Intelligence (IDSIA)}
  \city{Lugano}
  \country{Switzerland}}
\email{francesca.mangili@supsi.ch}
\orcid{0000-0002-3215-1028}

\author{Luca Maria Gambardella}
\affiliation{%
  \institution{Università della Svizzera italiana (USI), Dalle Molle Institute for Artificial Intelligence (IDSIA)}
  \city{Lugano}
  \country{Switzerland}}
\email{luca.gambardella@usi.ch}

\renewcommand{\shortauthors}{Assaf et al.}

\begin{abstract}
  This paper addresses the incorporation of problem decomposition skills as an important component of computational thinking (CT) in K-12 computer science (CS) education. Despite the growing integration of CS in schools, there is a lack of consensus on the precise definition of CT in general and decomposition in particular. While decomposition is commonly referred to as the starting point of (computational) problem-solving, algorithmic solution formulation often receives more attention in the classroom, while decomposition remains rather unexplored. This study presents ``CTSKills'', a web-based skill assessment tool developed to measure students' problem decomposition skills. With the data collected from 75 students {in grades 4-9}, this research aims to contribute to a baseline of students' decomposition proficiency in compulsory education. Furthermore, a thorough understanding of a given problem is becoming increasingly important with the advancement of {generative artificial intelligence (AI)} tools that can effectively support the process of formulating algorithms. This study highlights the importance of problem decomposition as a key skill in K-12 CS education to foster more adept problem solvers.
\end{abstract}





\maketitle

\section{Introduction}
Computer science (CS) has been increasingly introduced into K-12 education over the past decade. The term ``computational thinking'' (CT) was first introduced by Wing \cite{Wing2006} as a key competence in CS education. Sentance et al. \cite{Sentance23} informally paraphrase CT according to Wing \cite{Wing2006} as \textit{``the mental activity in formulating a problem to admit a computational solution. The solution can be carried out by a human or a machine. CT is not just about problem-solving but also about problem formulation.''}
Sentance et al. highlight the importance of problem formulation as a fundamental aspect of the problem-solving process. The introduction of the term CT has initiated an ongoing debate about its definition and the processes involved in designing activities to foster students' CT competencies. There is not yet an established and agreed-upon list of the most useful elements for K-12 CS education \cite{Rich2019,Dietz2019,Sjödahl2023}. 
Sentance et al. describe CT as a list of concepts such as \textit{logic and logical thinking, algorithms and algorithmic thinking, patterns and pattern recognition, abstraction and generalisation, evaluation, automation} and a list of practices: \textit{problem decomposition, creating computational artefacts, testing and debugging, iterative refinement (incremental development), collaboration and creativity (part of broader twenty-first-century skills).}

Consequently, there are several concepts and practices involved in CT. However, concepts such as algorithmic thinking have been given greater emphasis than others in daily classroom practices. A literature review on CT in education reveals that research tends to focus on learning outcomes and artefacts rather than the thinking process itself \cite{Tsai2021}. Decomposition is consistently identified in definitions of CT. However, it requires greater emphasis and remains an under-explored practice in CS education \cite{Rich2019,Dietz2019,Sjödahl2023}. Problem decomposition is frequently regarded as a starting point for problem-solving, encompassing computational problem-solving and many other fields. However, in the context of CS education, there is a lack of clarity regarding the definition and assessment of decomposition skills \cite{Rich2019,adorni2024_framework}. Consequently, there is no common decomposition proficiency baseline in compulsory education.

We believe that it is crucial in K-12 CS education research to address this issue and provide an instrument that contributes to assessing students' problem decomposition skills in CT. In this paper, we describe ``CTSkills'', a prototype of a decomposition skills assessment tool we have developed as a web application that can be easily used in schools. With this tool, we aim to contribute to CT research and provide a means of automatically collecting a large amount of data. This data could allow us to understand a baseline of decomposition skills to be expected according to a student's age. It can further provide a basis for further developing the underlying skills model. The CTSkills app has been designed for students across a wide age range. {Its initial prototype has been tested in a pilot study with 75 students in grades 4-9, and the results of this data collection are presented in this paper.} 
In light of the current situation, the following research questions emerge:
\begin{itemize}[noitemsep,nolistsep]
\item \textit{RQ1: How can problem decomposition be assessed in K-12 classrooms?} 
\item \textit{RQ2: What problem decomposition skills do students have across grades 4-9?}
\end{itemize}
This pilot study focuses on grades 4-9, but the overall aim is to expand the age range across grades K-12 with a more mature prototype of the assessment instrument. This pilot study aims to demonstrate the feasibility of assessing problem decomposition skills in the classroom and gain further insights for future work.  

\section{Related Work}
Problem decomposition is defined as the ability or process of breaking down problems into smaller sub-problems \cite{Rich2019,Kwon_Cheon_2019,Israeletal2023,Yadavetal2022,adorni2024_framework}. In more detail, Barr \& Stephenson \cite{BarrStephenson2011} describe decomposition as a core function to identify sub-problems and define the objects and methods needed in each decomposed task to solve a problem. According to Shute et al. \cite{shuteetal2017}, the smaller parts of the problem are functional elements that together represent the whole problem rather than random parts. However, there is no consensus regarding the cognitive abilities or the process of problem decomposition. Dietz et al. \cite{Dietz2019} aim to bridge the gap between cognitive development and CS education research. They looked into the basic capacities for problem decomposition in a physical/spatial domain in early childhood and found that even 4-year-olds can successfully evaluate the viability of decomposition plans. Sjödahl et al. \cite{Sjödahl2023} suggest that abstraction and decomposition are not two connected iterative practices but rather intertwined components of a process towards a solution. They studied students' programming in ScratchJr and identified abstraction and decomposition practices in their actions.     

P.J.Rich et al. \cite{Rich2019} analyse the decomposition processes across various STEM and non-STEM fields and developed a framework for decomposition in CT. This framework is in the context of CS education and \textit{``[...] may help educators to better prepare students to break down complex problems, as well as provide guidance for how decompositional ability might be measured''.}   P.J.Rich et al. further state: \textit{``before successful decomposition is performed, a problem or sub-problem is known only as a black box. The inputs, outputs, and other relationships between problems or sub-problems may be known, but the inner workings of the problem at hand are unknown.''}
The authors introduced three main categories of decomposition such as \textit{substantive decomposition}, \textit{relational decomposition}, and \textit{functional decomposition}. With these categories, the process of decomposing an overall problem can be better understood. That is, as an iterative process involving multiple steps that can be repeated until enough information is gathered to reconstruct the solution through patterns and algorithms. 

K.M.Rich et al. \cite{RichKathryn2018} introduced decomposition learning trajectories for K-8 CT. Based on an extensive literature review, they identified 63 learning goals that represent decomposition understanding and practices and synthesised them into 13 consensus goals such as ``systems are made up of smaller parts'', ``code is reusable'', or ``defining procedures allows for easy reuse of code within programs''. Furthermore, they established relationships between these consensus goals and placed them into learning trajectories to facilitate the design of assessments and differentiation. In relational decomposition, each component or sub-component of the problem may relate to other parts and sub-parts. Those relations could be based on time, sequence, location, dependence, etc.   

Regarding the assessment of decomposition skills, studies often focus on assessing general aspects of CT. 
Tang et al. \cite{Tangetal} provided an overview of CT assessments in surveys, portfolios, traditional assessments, and interviews. 
The ``CT Test'' by Roman-Gonzalez et al. \cite{RomanGonzalez2017} or the ``CT Scale'' by Korkmaz et al. \cite{Korkmaz2017} do not contain detailed information on the evaluation of decomposition. 
To date, decomposition has been primarily determined and assessed based on learning outcomes using artefacts or program components \cite{Liu2021,Tsai2021}.
One example is the Dr. Scratch tool \cite{MorenoLeon2015}, which automatically analyses scratch artefacts and evaluates decomposition based on blocks used. Kwon and Cheon \cite{Kwon_Cheon_2019} also focused on learning outcomes in the form of Scratch artefacts, while Israel et al. \cite{Israeletal2023} used an item test to assess decomposition.

\section{Method}
\subsection{Research Setting and Approach}
This study was conducted as part of an ongoing research project to define and assess CT in the context of K-12 CS education. Previously, our focus was on algorithmic thinking and its assessment through unplugged activities in K-12 \cite{piatti_2022}. Recently, a larger study compared the unplugged activity with an automated version implemented as a native app running on mobile devices \cite{adorni_chbr,adorni_pilot2023,adorni_softwarex}. In the context of this research project, we are further focusing on assessing problem decomposition skills. Building on our previous experience of automatically assessing algorithmic thinking through an app, we have adopted a similar quantitative approach for assessing decomposition skills. The automated data collection allows us to reach a large number of students more easily.

We use the aforementioned theoretical framework for decomposition in CT by P.J.Rich et al. \cite{Rich2019} to design the assessment instrument, where a problem is referred to as a black box. In a CS education context, a ``problem'' might be an interactive game that a student wants to develop in a Scratch programming environment. The student has an idea of the input and output of the game and might have developed a story of what should happen in the game but does not yet have an idea of the programming code. To implement the game, the student needs to \textit{``[...] unpack the problem and separate it into multiple sub-problems. [...] the sub-problems all begin as black boxes, with their inner structures unknown to the problem solver.''}\cite{Rich2019}. Let's assume the student wants to implement a simple interactive game where a user collects apples from a tree and places them in a basket, increasing a score. Apples can also be dropped on the ground and become spoiled. To implement this game in a Scratch programming environment, substantive decomposition is a starting point to solve this problem. 
P.J.Rich et al. referred to axis selection in the process of substantive decomposition as dependent on the problem statement and context. In the context of the interactive game to be implemented by the student, the axis might be to break down the game scenery by its component parts, separating connected or interacting objects by their shape, look, or behaviour. This process involves the abstraction of the relevant objects in the scenery. This could be the apples, the basket, and the score. The relational decomposition is performed by assigning a relation between two components or sub-components of a problem. To achieve this, the relations between the objects in the game must be defined, such as ``apple relates to basket'', ``apple in basket relates to score'', ``apple relates to spoiled apple'' etc. One approach is to assign properties and behaviours to the objects, such as ``apple collides with basket'' and ``apple collides with ground'' to create a model of the system.   

Once the problem is decomposed into formally described sub-problems, they can be sorted, grouped, categorised or decomposed further. Similar objects can be abstracted and generalised into classes, functions can be created, etc. The student will understand that the properties and behaviours of an apple might be similar to another fruit on another tree in the game. An advanced skilled student will create modular programming code with reusable functions, leading to the learning goals such as ``code is reusable'', or ``defining procedures allows for easy reuse of code within programs'' according to K.M.Rich et al.'s decomposition learning trajectories \cite{RichKathryn2018}. Creating modular, reusable, and dynamic code that avoids repetitive, hard-coded and static procedures is an indicator of a skilled learner with decomposition as the starting point to problem-solving.      

\subsection{Research Instrument}
\begin{figure}[htb]
  \centering
  \begin{subfigure}[b]{0.33\textwidth}
    \centering
    \includegraphics[width=\linewidth]{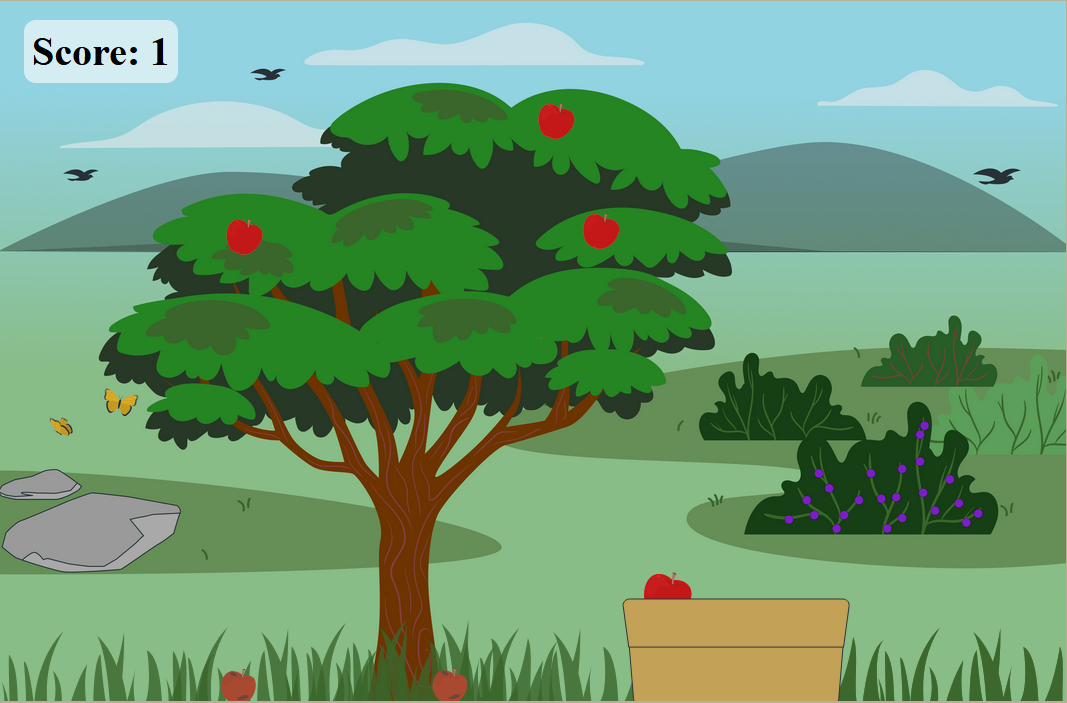}
    \caption{Level1}
    \label{fig:ctskills_level1}
  \end{subfigure}
  \hfill
  \begin{subfigure}[b]{0.33\textwidth}
    \centering
    \includegraphics[width=\linewidth]{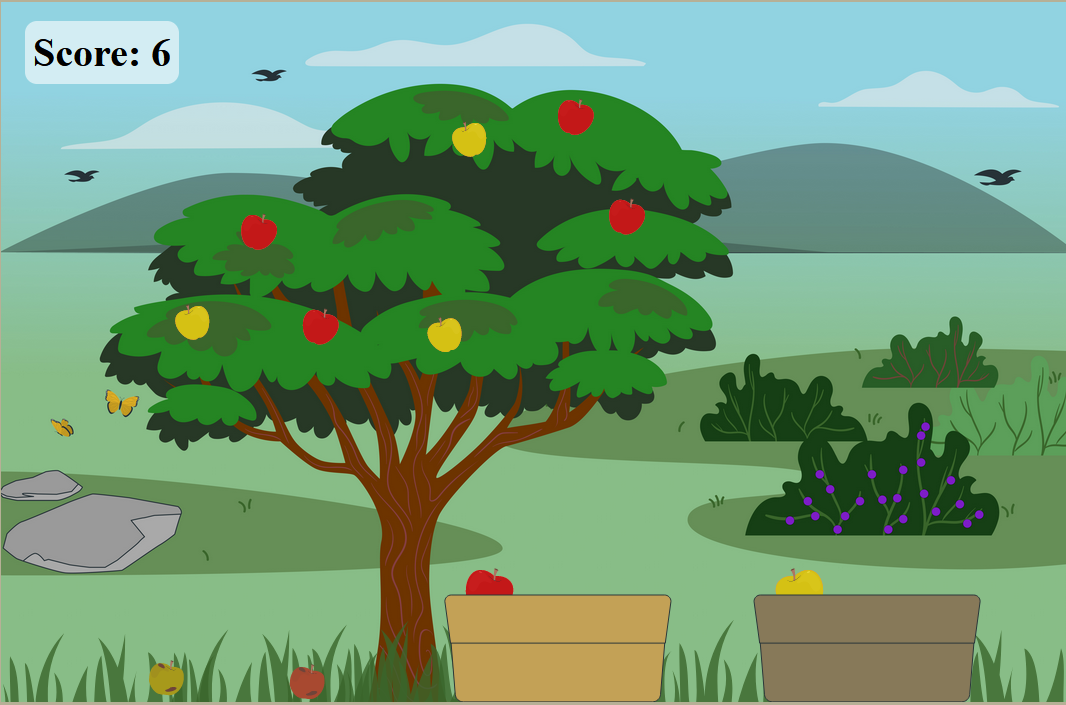}
    \caption{Level2}
    \label{fig:ctskills_level2}
  \end{subfigure}
    \hfill
  \begin{subfigure}[b]{0.33\textwidth}
    \centering
    \includegraphics[width=\linewidth]{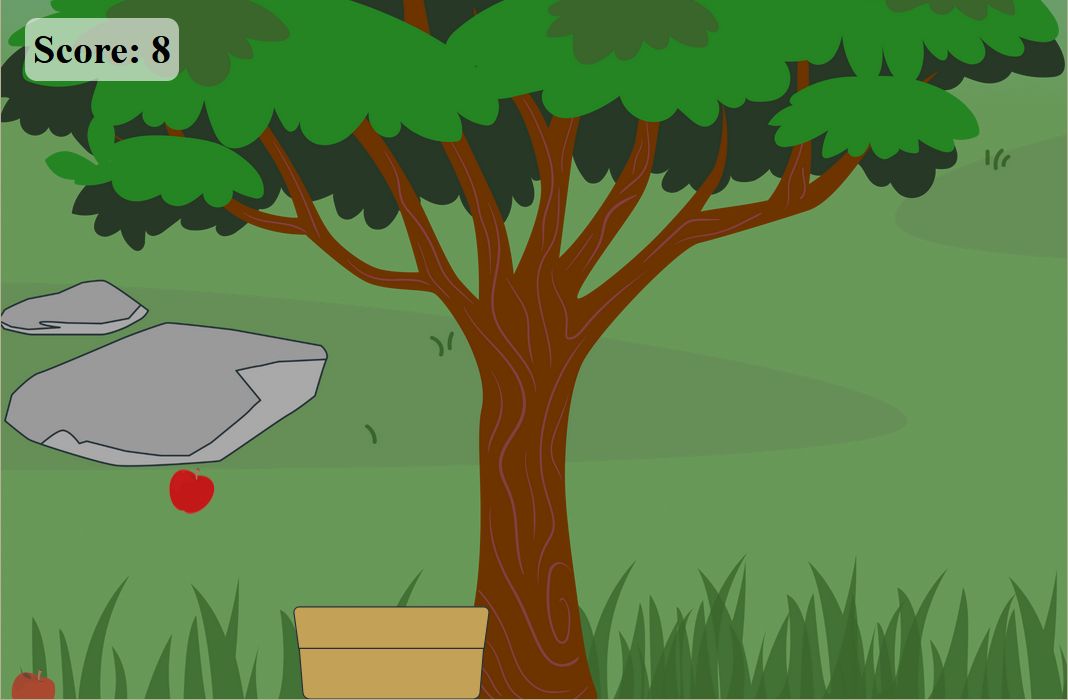}
    \caption{Level3}
    \label{fig:ctskills_level3}
  \end{subfigure}
  \caption{Screenshots from Level 1, Level 2, and Level 3 of the CTSkills app.}
  \label{fig:ctskills_allLevels}
  \Description{}
\end{figure}

We developed a web application called ``CTSkills'' using Javascript, HTML, and CSS that runs on any browser. It has been optimised for mobile devices such as tablets for the school context. During the development phase, the app underwent usability tests with seven children aged 6-13. These tests focused on key usability aspects, including the intuitiveness of the graphical user interface and the clarity of the app labels. The application comprises three levels of a simple interactive game, referred to as ``scenery'', illustrated in Fig.~\ref{fig:ctskills_allLevels}. 

The first level's scenery features a tree with red apples, a basket on the ground, and patches of grass in a landscape with a number of different objects, including rocks, bushes, clouds, butterflies and birds (Fig.~\ref{fig:ctskills_level1}). Additionally, a score is set initially to zero. The only objects that can be dragged in this scenery are the apples on the tree. Users can drag them into the basket or drop them on the grass. Each time an apple is placed in the basket, the score increments. On the other hand, when the apple is dropped on the ground, it becomes spoiled and cannot be dragged again. Whenever an apple is dragged elsewhere in the scenery, it returns to its original position on the tree. Once all the apples have been placed in the basket or dropped on the grass, the app will transition to a question screen.

\begin{figure}[htb]
  \centering
  \begin{subfigure}[b]{0.48\textwidth}
    \centering
    \includegraphics[width=\linewidth]{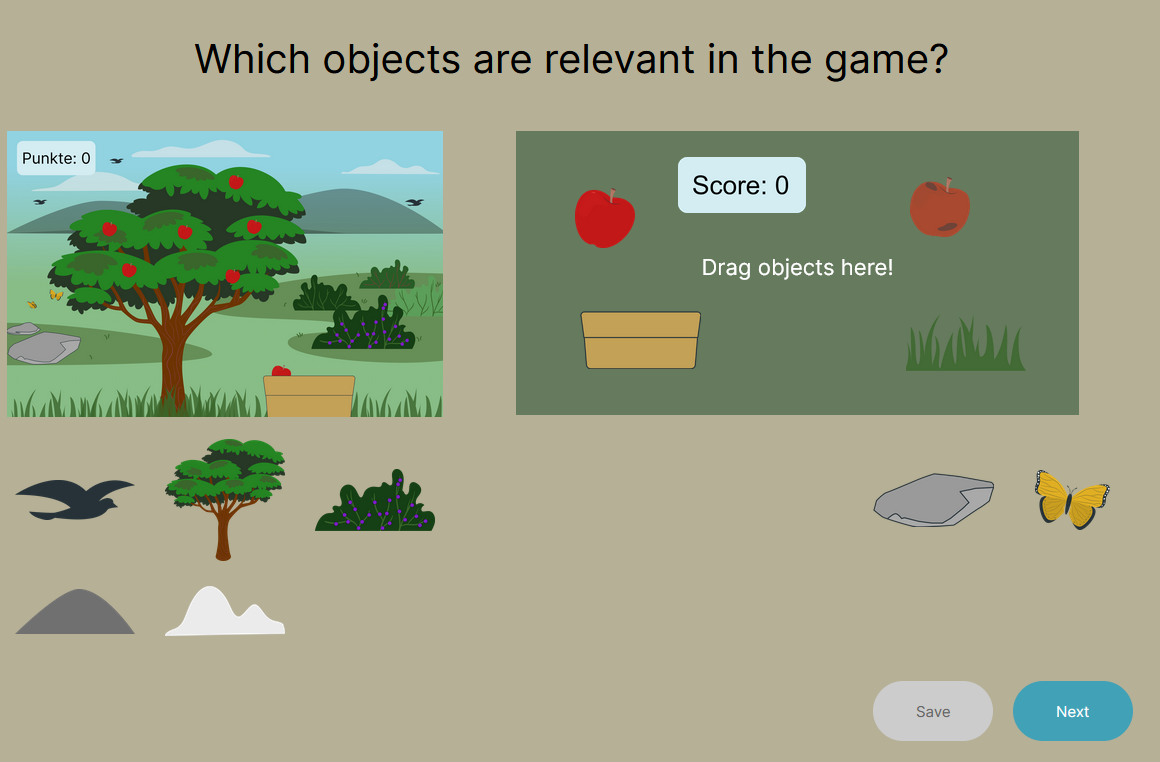}
    \caption{Level 1, Q1: Substantive decomposition and abstraction}
    \label{fig:ctskills_level1_Q1}
  \end{subfigure}
  \hfill
  \begin{subfigure}[b]{0.48\textwidth}
    \centering
    \includegraphics[width=\linewidth]{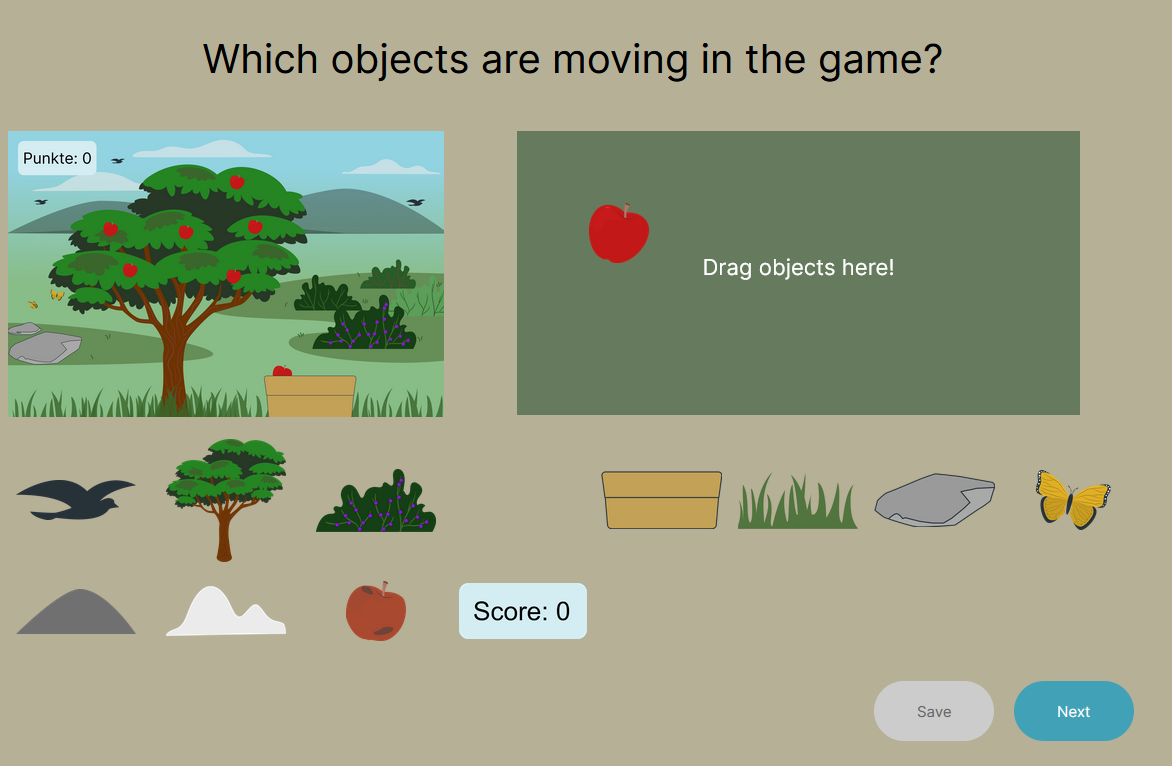}
    \caption{Level 1, Q2: Assigning properties to objects}
    \label{fig:ctskills_level1_Q2}
  \end{subfigure}
  \\
  \begin{subfigure}[b]{0.48\textwidth}
    \centering
    \includegraphics[width=\linewidth]{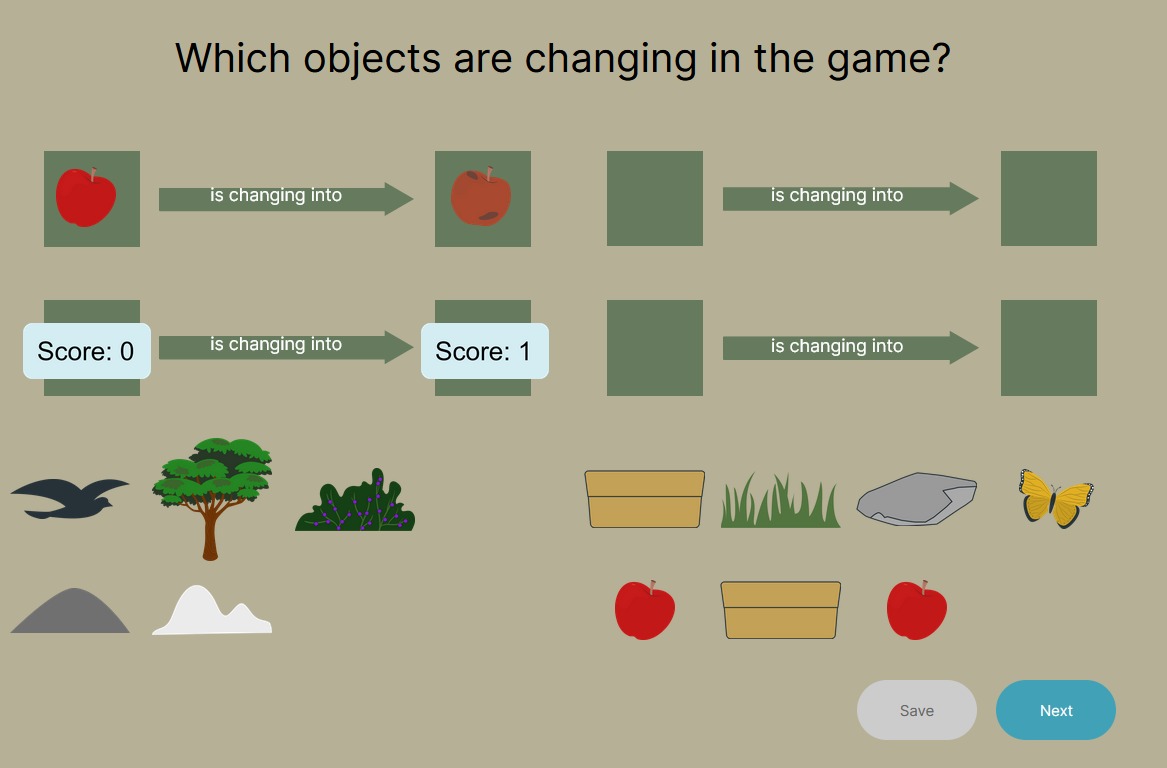}
    \caption{Level 1, Q3: Relational decomposition}
    \label{fig:ctskills_level1_Q3}
  \end{subfigure}
  \hfill
  \begin{subfigure}[b]{0.48\textwidth}
    \centering
    \includegraphics[width=\linewidth]{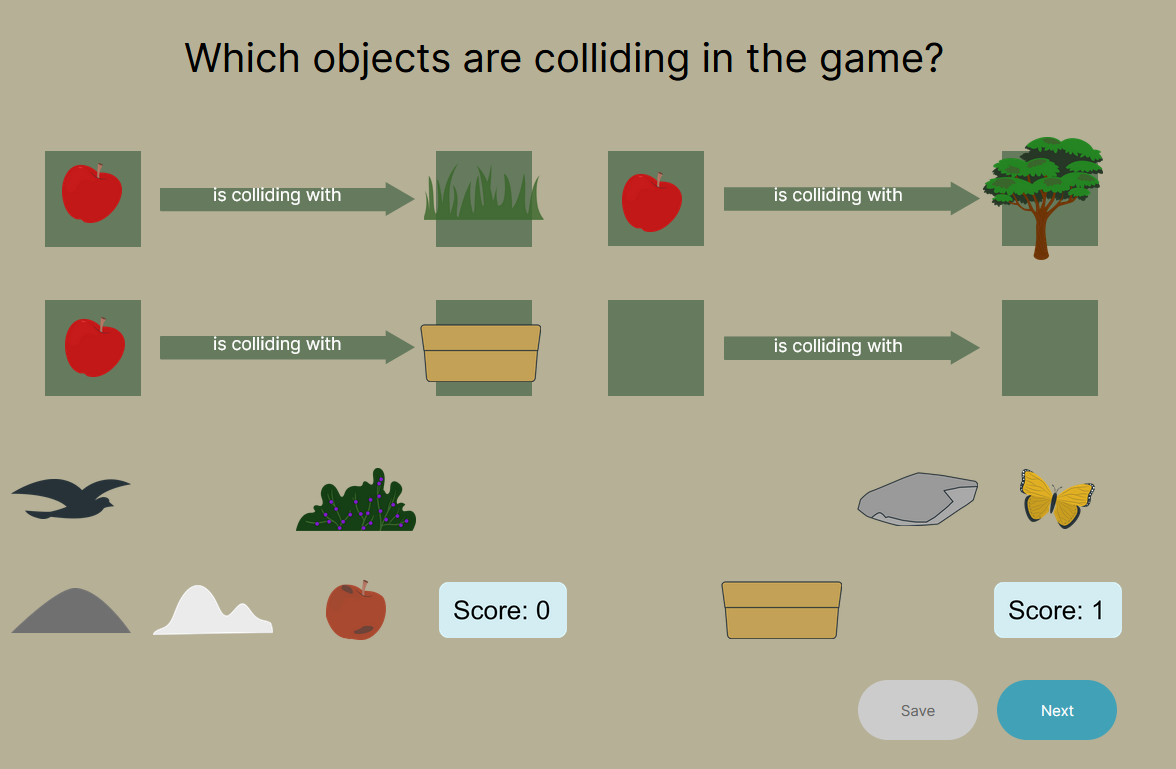}
    \caption{Level 1, Q4: Relational decomposition}
    \label{fig:ctskills_level1_Q4}
  \end{subfigure}
  \caption{Screenshots from Question 1 (Q1) to Question 4 (Q4) of Level 1 of the CTSKills app.}
  \label{fig:ctskills_level1_allQuestions}
  \Description{}
\end{figure}

In question screen (Q1), the user is asked to drag all relevant objects in this scenery into the green solution field. These may include apples, spoiled apples, a basket, a score, and grass (Fig.~\ref{fig:ctskills_level1_Q1}). It is evident that this leaves room for interpretation. One might argue that the tree is as relevant as the apple. From a game developer's perspective, the tree is simply part of the background without interacting with objects. It seems important since the apples are returned to their original position on the tree when dragged elsewhere in the scenery, but their home coordinates have no relation to the tree object itself. The other objects in the scenery, such as birds, rocks, and bushes, are similarly irrelevant in this game. This question refers to substantive decomposition according to the framework by P.J.Rich et al. \cite{Rich2019} and the ability to separate relevant from irrelevant objects (abstraction). 
Upon clicking the ``Next'' button, the user is prompted to determine which objects are moving within the given scenery. Question screen (Q2) relates to assigning properties to objects (Fig.~\ref{fig:ctskills_level1_Q2}). Here, the correct answer would be the apple.     
The third question (Q3) concerns decomposing into relational categories or the relationships between objects. The user must place the different objects on either side of an arrow labelled ``is changing into''. Correct relationships include the red apple that transforms into a spoiled apple and the incremented score (Fig.~\ref{fig:ctskills_level1_Q3}). 
The final question (Q4) also refers to relational decomposition. The user is required to identify the objects that collide in the scenery. The correct answers include the red apple colliding with the grass or basket. Optional but not considered incorrect if the apple is seen as colliding with the tree upon its return to the home position. (Fig.~\ref{fig:ctskills_level1_Q4}).   

The app then proceeds to the next scenery (Level 2), which is a tree with two coloured apples, red and yellow (Fig.~\ref{fig:ctskills_level2}). Two baskets are placed on the ground, indicating which basket each apple colour belongs to. The game works in a similar way as Level 1, with both apple colours able to be dropped on the ground and turned into spoiled apples. The correct apple colour must be placed in the corresponding basket; the incorrect colour is rejected, and the apple is returned to its original position on the tree. 
In the third scenery (level 3), there is a slight modification: the view is zoomed into the trunk area of the tree, and red apples and dark and light-coloured leaves are falling on the ground (Fig.~\ref{fig:ctskills_level3}). The basket is now horizontally draggable with which the apples can be caught and which causes the score to increment. Conversely, leaves do not contribute to the score, and any apples that fail to be caught and instead touch the grass are transformed into spoiled apples. 
Each scenery is followed by the same four questions Q1-Q4 as the previous level but with an altered selection of objects that reflect the new scenery. For instance, in Level 2, a yellow basket is introduced, along with a yellow apple and a spoiled yellow apple. Similarly, in Level 3, the selection includes dark and light-coloured leaves as objects. 

\begin{figure}[htb]
  \centering
  \begin{subfigure}[b]{0.325\textwidth}
    \centering
    \includegraphics[height=3.25cm]{pics/CTSKills_level1_Q4.PNG}
    \caption{Level 1, Q4}
    \label{fig:ctskills_level1_q4}
  \end{subfigure}
  \hspace{.05cm}
  \begin{subfigure}[b]{0.325\textwidth}
    \centering
    \includegraphics[height=3.25cm]{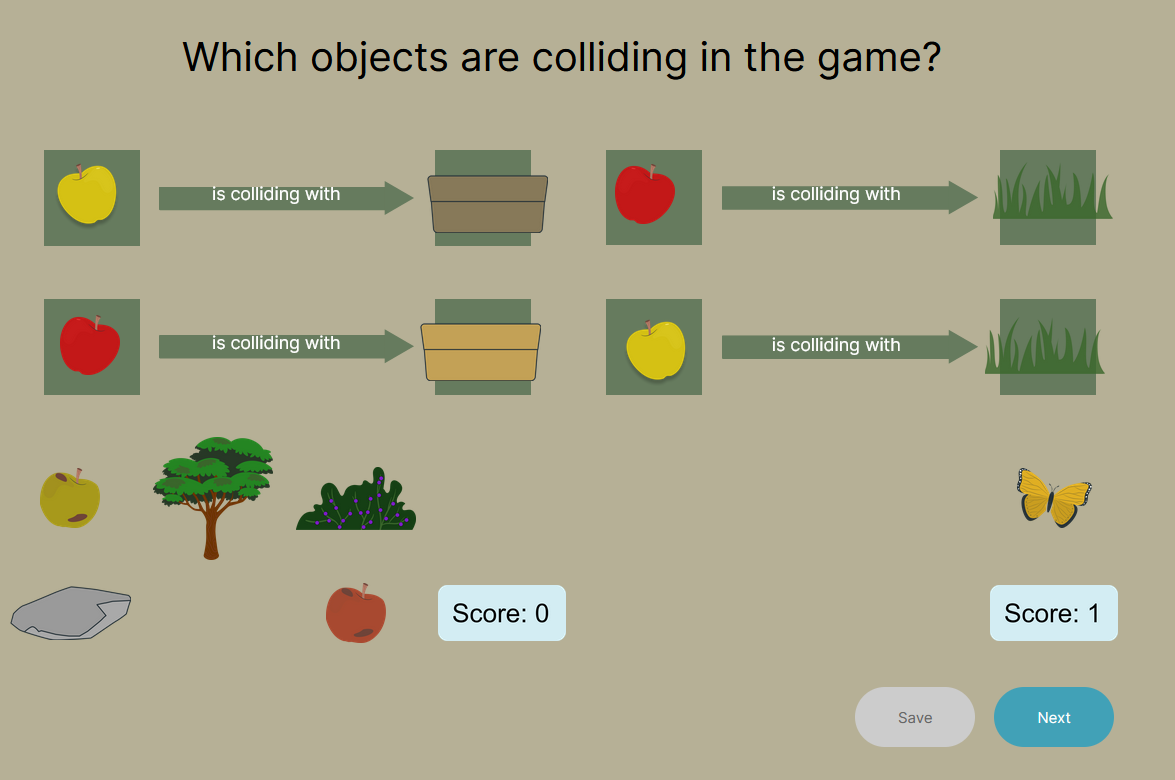}
    \caption{Level 2, Q4}
    \label{fig:ctskills_level2_q4}
  \end{subfigure}
  \hspace{.05cm}
  \begin{subfigure}[b]{0.32\textwidth}
    \centering
    \includegraphics[height=3.25cm]{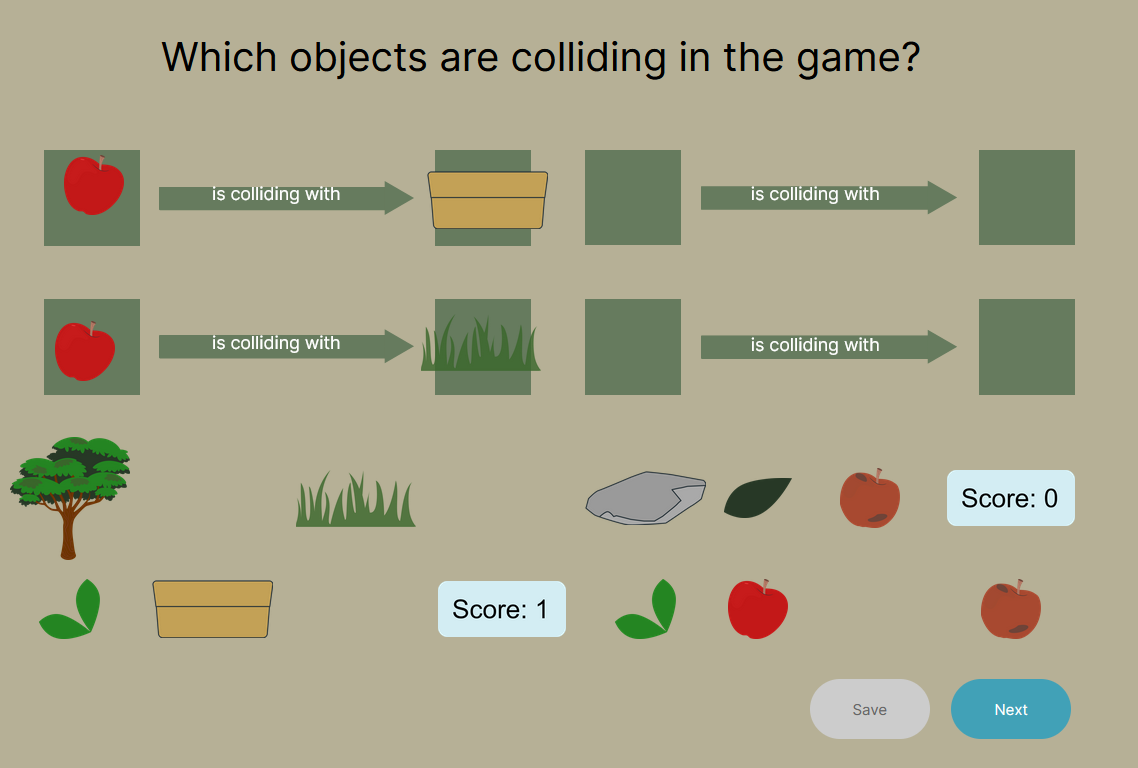}
    \caption{Level 3, Q4}
    \label{fig:ctskills_level3_q4}
  \end{subfigure}
  \caption{Pattern recognition, abstraction and generalisation to make code reusable.}
  \label{fig:ctskills_questions4}
  \Description{}
\end{figure}

The levels were designed such that objects are repeated in different contexts, sizes, or shapes. The student should realise that a yellow apple in Level 2 does not differ much from the red apple in size, shape, and behaviour. It can be abstracted and generalised into a class ``apple'' with common properties and functions. Also, a falling apple in level 3 that collides with the basket leads to an increased score, not only an apple that was manually dragged into the basket. 
A falling leaf and a falling apple exhibit similar behaviours. By looking at the solutions of the relational decomposition question 4 (Q4) for all three levels, the student might realise through pattern recognition (another CT skill) the repeating relations within one level and across several levels (Fig.~\ref{fig:ctskills_questions4}).
This initial prototype of the CTSkills app incorporates the substantive and relational decomposition of similar scenarios of different complexities. This is the starting point for more sophisticated skills, such as decomposing into functional categories for more advanced skilled learners. 

\subsection{Assessment}
The scoring methodology assesses students' accuracy in identifying target items and avoiding non-target items across various questions and levels.
Each combination of questions and levels specifies a predefined set of targets $X$ and non-targets $Y$, with distinctions in the type of targets involved: questions Q1 and Q2 focus on individual target elements, while Q3 and Q4 involve pairs of target elements.
%
The score calculation is based on the selected sets $S_X \subseteq X$  (target items or pairs) and $S_Y \subseteq Y$ (non-target items or pairs):
\[
\text{score} = |S_X| - (|X|-|S_X|) - |S_Y| \mbox{.}
\]
This formula rewards correct selection of targets, penalises missed targets and selection of non-targets.
Scores are assigned only if the student attempted the exercise.

A rescaling method is employed to standardise scores across assessments. It begins by computing the minimum achievable score $\text{min\_score}=-|X|+|Y|$, and then transforming the initial score into a standardised range from 0 to 5:
\[
\text{rescaled\_score} = \text{max\_scaled} \times \frac{\text{score} - \text{min\_score}}{|X| - \text{min\_score}} \mbox{,}
\]
where \(\text{max\_scaled}\) represents the upper boundary of the rescaled score range.

Post-assessment, average scores are computed across the 4 questions and their respective levels. These averages yield a final aggregate score per student, offering a compact performance indicator for the study.

\subsection{Participants}

We conducted a pilot study in seven classes in Spring 2024, involving 75 students aged 10-17. 
The detailed breakdown of students demographics is presented in Table~\ref{tab:participants}.
Except for the students in the first session, all classes had previous exposure to Scratch programming. Students in grades 4-6 attended standard primary school, while those in grades 7-9 were enrolled in secondary school.
\begin{table*}[ht]
\footnotesize
\centering
\caption{{Demographic Analysis of Students in the Pilot Study.} This table provides an overview of student demographics by session, including school type, grade, age range (mean and standard deviation), and sex distribution (number of female and male students).}
\label{tab:participants}
\begin{tabular}{ccccccc}
\hline
 \textbf{Session}   & \textbf{School type}   &   \textbf{Grade} & \textbf{Age range}&   \textbf{Female} &   \textbf{Male} &   \textbf{Total} \\
\hline
 1    & Primary School & 4 & 10 - 10 years ($\mu$ = 10.0 $\pm$ 0.0) &   4 & 5 &  9 \\
 2    & Primary School & 5 & 10 - 12 years ($\mu$ = 10.9 $\pm$ 0.5) &   9 & 7 & 16 \\
 3    & Primary School & 6 & 11 - 12 years ($\mu$ = 11.9 $\pm$ 0.4) &   3 & 5 &  8 \\
 4    & Secondary School& 7 & 13 - 14 years ($\mu$ = 13.2 $\pm$ 0.5) &   1 & 3 &  4 \\
 5    & Secondary School&8 & 13 - 15 years ($\mu$ = 13.9 $\pm$ 0.6) &   6 & 8 & 14 \\
 6    & Secondary School&9 & 15 - 15 years ($\mu$ = 15.0 $\pm$ 0.0) &   2 & 2 &  4 \\
 7    & Secondary School&9 & 14 - 17 years ($\mu$ = 15.3 $\pm$ 0.7) &  11 & 9 & 20 \\
 \cmidrule(lr){1-1}\cmidrule(lr){2-2}\cmidrule(lr){3-3}\cmidrule(lr){4-4}\cmidrule(lr){5-5}\cmidrule(lr){6-6}\cmidrule(lr){7-7}
 \textbf{{{Total}}} &   &    & 10 - 17 years ($\mu$ = 13.0 $\pm$ 2.1) &  36 &39 & 75 \\
\hline
\end{tabular}\end{table*}

The study was set up in a classroom environment where students used the CTSkills app individually. Each student had access to a school-provided iPad and used the Safari browser to interact with the app. The session lasted approximately 20 minutes, during which students independently navigated through the app's levels and questions. Assistance was provided when necessary, especially for clarifications on terms like  ``collision'' or ``change''.

Data collection involved capturing automatically generated screenshots and saving data to a JSON file, which recorded user responses, timestamps, and object movements. 
An initial screen presented at the beginning of the activity collected demographic information about participants, including age, grade, gender, and language. No sensitive personal information was recorded beyond these details.

All collected data was securely stored on a file server managed by the authors. After data collection, all information and browser history were deleted from the iPads to ensure student privacy; teachers could not access individual student results. The authors can access the data stored on the secure server and download it as needed.


Informed consent was obtained from all participants and their guardians prior to the study. They were informed about the purpose of the study, the procedures involved, and their right to withdraw at any time without penalty.

Participants' data were anonymised to ensure privacy. Personal identifiers were removed, and data were stored securely on encrypted servers. Only authorised personnel had access to the data.

This study adhered to ethical standards and received approval from a committee. Participants and, for those under 12, their parents or legal guardians provided informed consent; those over 12 also gave assent. Following local and international guidelines, data was handled confidentially, with pseudonymisation for participant protection.


\section{Results}

In this paper, we describe an excerpt of all the data collected to illustrate the idea of the assessment tool. 




\begin{figure}[htb]
  \centering
  \includegraphics[width=\linewidth]{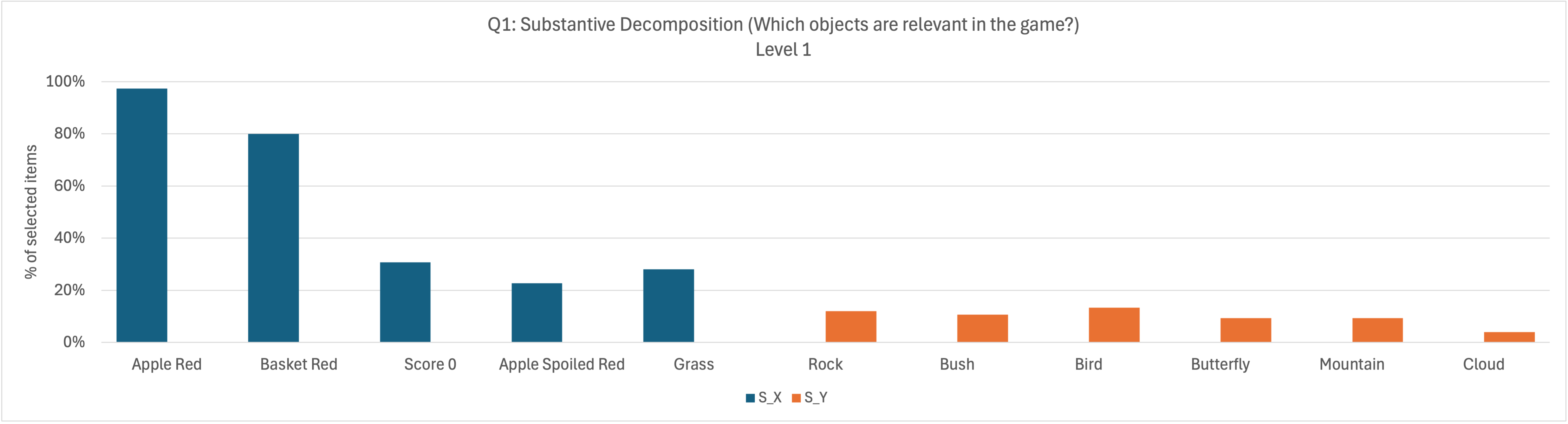}
  \Description{}
  \caption{{Percentage of Times Each Item Was Selected as Relevant for Question 1 - Level 1.}
  This chart illustrates the percentage of times each item was selected as relevant, differentiating between correctly selected targets $S_X$ (in blue) and incorrectly selected non-targets $S_Y$ (in orange).
  }
  \label{fig:ctskills_Q1_L1_results}
\end{figure}
Fig.~\ref{fig:ctskills_Q1_L1_results} showcases the results specific to Question 1 at Level 1, showing students' ability to identify targets like red apples and the basket. Other relevant objects received less recognition, indicating varying student proficiency levels across different items.
%
Performance improved notably at Level 2, showing strong identification of the new items introduced, although some items declined by Level 3 (Figures from this analysis are available in Supplementary material). 

\begin{figure}[htb]
  \centering
  \includegraphics[width=\linewidth]{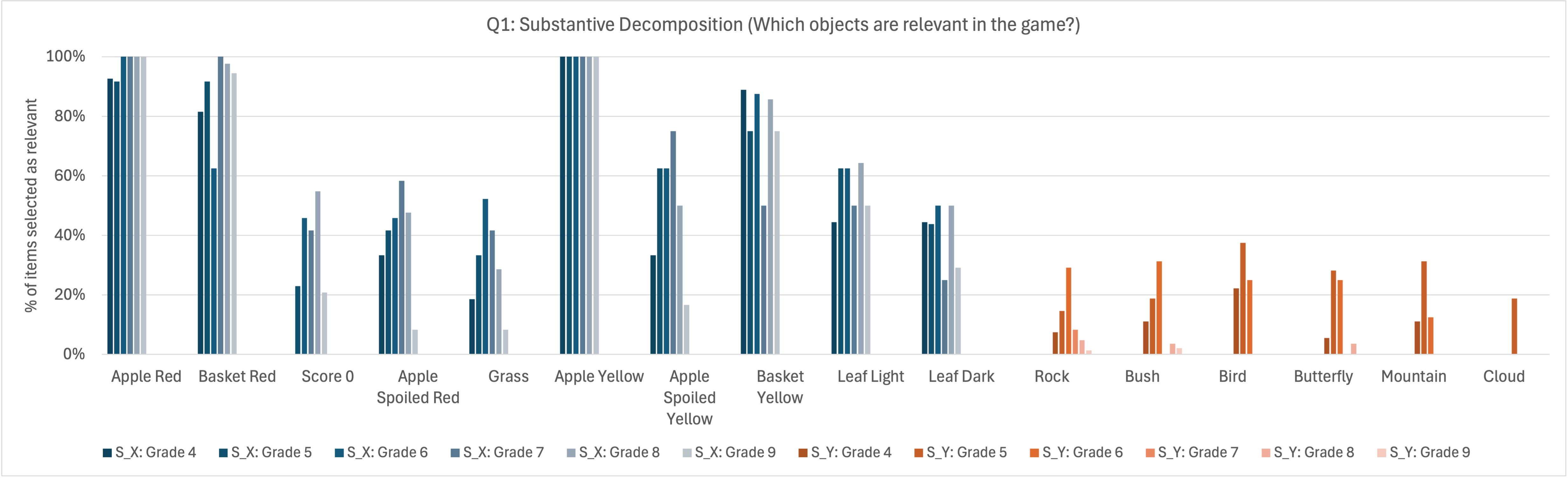}
  \Description{}
  \caption{{Percentage of Times Each Item Was Selected as Relevant for Question Q1 across School Grades.}
  This chart illustrates the percentage of times each item was selected as relevant, differentiating between correctly selected targets $S_X$ (in different shades of blue for each school grade) and incorrectly selected non-targets $S_Y$ (in different shades of orange for each school grade).
  }
  \label{fig:ctskills_Q1_results_by_grade}
\end{figure}
Fig.~\ref{fig:ctskills_Q1_results_by_grade} presents cumulative results across all levels and grades, illustrating a consistent trend of increasing correct selections $(S_X)$ as students progress through grades.
Notably, items like red apples and baskets maintained high selection rates from grade 8 to 11, while other items showed more variability, peaking in middle grades and then dropping in higher grades.
In contrast, the percentage of non-target elements incorrectly identified as targets $(S_Y)$ initially increased, peaking around grade 8, before decreasing in higher grades. 
This trend suggests that younger students struggle more with distinguishing non-target elements, but this ability improves as they advance in grade level.

\begin{figure}[htb]
  \centering
  \includegraphics[width=\linewidth]{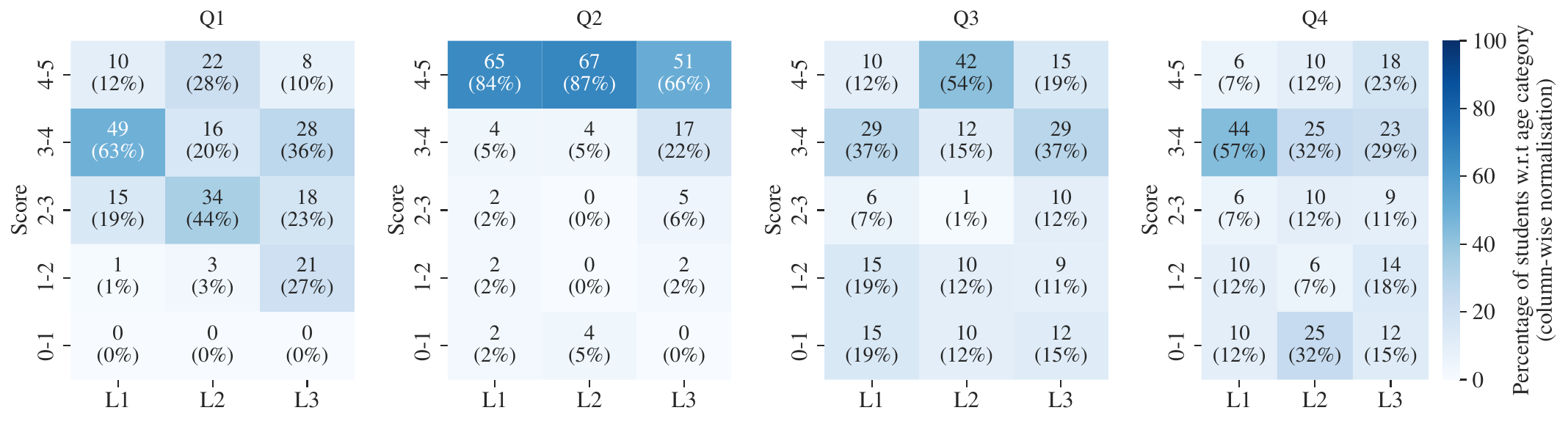}
  \Description{}
  \caption{{Percentage Distribution of Student Scores by Question and Level.}
  This graph illustrates the percentage distribution of student scores across four questions (Q1, Q2, Q3, Q4) and three difficulty levels (L1, L2, L3). 
  Each subplot represents a question, with difficulty levels indicated on the x-axis and score categories on the y-axis. 
  Annotations within each cell display the number of students achieving a specific score for that level, followed by the percentage relative to the total number of students. 
  }
  \label{fig:score_by_question}
\end{figure}
Shifting focus to overall scores across all questions, Fig.~\ref{fig:score_by_question} illustrates significant variations in score distributions among questions.
Q2 stands out with higher scores, potentially potentially indicating different difficulty levels or student proficiency.
Variability in performance across levels is evident, with Q2 and Q3 showing more consistency than Q1 and Q4.
Additionally, scores generally decrease with increasing difficulty levels across all questions, reflecting the challenge of higher-level tasks. 

\begin{figure}[htb]
  \centering
  \includegraphics[width=\linewidth]{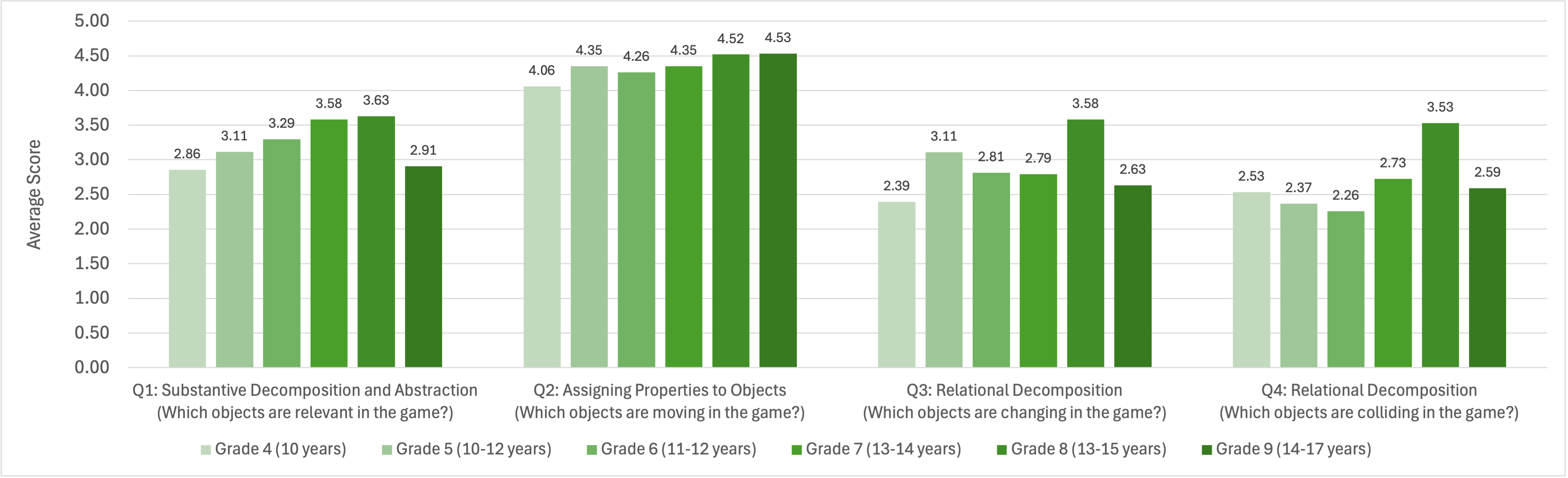}
  \Description{Score per grade.}
  \caption{Average Score across School Grades for Questions Q1-Q4 on Levels 1-3.}
  \label{fig:ctskills_averageScore_allGrades}
\end{figure}
Finally, Fig.~\ref{fig:ctskills_averageScore_allGrades} summarises the average scores across school grades for each decomposition question category, highlighting an overall improvement in decomposition skills with age, particularly in tasks involving substantive decomposition and abstraction (Q1) and assigning properties to objects (Q2), that appear easier than abstracting objects and establishing relations (Q3 and Q4).



Transitioning from the earlier examination of frequency distributions and performance patterns, our statistical analysis through ANOVA \cite{James2013,chambers1992,bartlett1937properties,Hastie2001}, Chi-Square Tests of Independence \cite{moore_introduction_1989,Yates1934,Cochran1954}, post-hoc tests using Tukey's HSD \cite{Tukey1949,moore_introduction_1989,Yates1934,Cochran1954,Sthle1989}, and Linear Mixed-Effect Models (LMMs) \cite{hox2017multilevel,raudenbush2002hierarchical,bartlett1937properties,chambers1992,Kuznetsova2017}
 provides deeper insights into the factors influencing score variations across grades, gender, and questions.
Significant findings include varying scores across different grades and questions. Both tests reveal significant effects of grades (ANOVA: [$F(5, 294) = 3.016, p = 0.0113$], Chi-Square: [$\chi^2 = 705.96, p = 0.0063$]) and questions (ANOVA: [$F(3, 296) = 41.5, p < 2e-16$], Chi-Square: [$\chi^2 = 625.12, p < 2e-15$]), while neither ANOVA $(F(1, 298) = 0.427, p = 0.514)$ nor Chi-Square $(\chi^2 = 116.32, p = 0.6521$ show significant effects of gender on scores.
Additionally, for grades and questions, notable differences were highlighted by post-hoc tests. 
Grade 8 had significantly higher scores than Grade 4 $(\text{MD}=0.8527, p = 0.0125)$, but unexpectedly, Grade 9 had significantly lower scores than Grade 8 $(\text{MD}=-0.6500, p = 0.0175)$. 
Regarding questions,  Question 2 had significantly higher scores than Question 1 $(\text{MD}=1.2397, p < 0.0001)$, Question 3 had significantly lower scores than Question 2 $(\text{MD}=-1.4865, p < 0.0001)$ and Question 4 had significantly lower scores than Question 1 $(\text{MD}=-0.4725, p = 0.0266)$ and  Question 2 $(\text{MD}=-1.7123, p < 0.0001)$.
Furthermore, LMMs elucidate the impact of questions and student-level factors on score variability. 
The models implemented included random intercepts for questions, grades, and students and gender as a fixed effect. 
Initial analysis showed that gender had a non-significant impact on scores $(p = 0.429)$, while the random intercepts for all factors are all significant $(p < 0.05)$, supporting their inclusion in the model.
Insights from this model underscored significant score variability attributed to questions $(\text{Variance}=0.573, \text{SD}=0.757)$ and notable student variability $(\text{Variance}=0.162, \text{SD}=0.402)$. Grade contributed minimally $(\text{Variance}=0.061, \text{SD}=0.247)$ to score variability.

\section{Discussion and Conclusion}
This pilot study focused on assessing substantive and relational decomposition skills among students in grades 4 to 9 using the ``CTSkills'' app. Built upon relevant theoretical frameworks, the app proved effective in automating data collection across a wide age range in classroom settings (RQ1). Although the study was restricted to grades 4 to 9 due to reading requirements, insights from usability testing with 6-year-olds after task clarification contributed valuable findings on abstraction and decomposition skills across all grade levels (RQ2).
%
%
Key insights from our study include a consistent improvement in task performance as students progress through grades, reflecting cognitive development in CT skills. 
Notably, Grade 9 exhibited unexpectedly lower performance than Grade 8, likely due to the increased complexity of secondary-level computational concepts and heightened academic expectations during this transitional period. 
Future research could explore interventions to support Grade 9 students in mastering these skills.
%
%
%
Importantly, there were no significant gender differences in decomposition scores, suggesting the equitable distribution of these skills across genders and underscoring the inclusivity of CT education.



\subsection{Limitations}
Despite these insights, our study faces limitations. 
Firstly, the pre-abstraction of objects within the app's menu may have influenced how students approached task identification and classification. 
Secondly, variations in language comprehension among students could have impacted their understanding of task instructions, potentially affecting performance consistency. 
Additionally, assessing creative solutions, such as grouping apples and baskets together, posed challenges in quantification and evaluation. 
Moreover, our reliance on cross-sectional data limits establishing causal relationships or tracking longitudinal developments in decomposition skills. 
Factors like individual student motivation, prior knowledge, and classroom dynamics were not fully captured, potentially influencing performance outcomes.


\subsection{Future Work}
Moving forward, this pilot study lays the groundwork for enhancing the CTSkills app's usability and task difficulty assessment among its intended users. 
Based on initial findings, future app revisions will include language options and incorporate more complex tasks like functional decomposition across various grade levels. 
Ongoing data collection in classrooms will further refine the app's functionality.
%
%
Future iterations of the app will aim to automatically assess students' skills and dynamically adjust task difficulty based on performance metrics derived from this research. 
Refined score assessment methods will be considered. 
Currently, we use an average score approach, but our findings indicate considerable performance differences among questions. Weighting scores could be beneficial if certain questions or levels are deemed more critical or representative of learning objectives. Alternatively, equal weighting may be chosen to maintain simplicity and fairness across all assessment components.
Longitudinal research designs tracking individual student progress over multiple academic years will deepen our understanding of decomposition skill development. Exploring additional factors such as teacher effectiveness and classroom environment through qualitative methods will provide richer insights into the educational contexts influencing student learning in CT skills.
Addressing these considerations in future studies will advance our understanding of effective educational interventions and instructional practices to support diverse student populations in developing decomposition and CT skills.

\section*{Fundings}
 This research was funded by the Swiss National Science Foundation (SNSF) under the National Research Program 77 (NRP-77) Digital Transformation (project number 407740\_187246).
 
\section*{Competing Interests}
 The authors declare that they have no conflict of interest.

\section*{Author Contributions}
\textbf{Dorit Assaf}: Conceptualization, Methodology, Software, Validation, Investigation, Data curation, Writing - original draft \& review \& editing, Visualization, Supervision, Project administration, Funding acquisition.\\
\textbf{Giorgia Adorni}: Conceptualization, Methodology, Validation, Formal analysis, Data curation, Writing - original draft \& review \& editing, Visualization. \\
\textbf{Elia Lutz}: Conceptualization, Methodology, Validation, Formal analysis, Investigation, Data curation, Writing - original draft \& review \& editing, Visualization. \\
\textbf{Lucio Negrini}: Conceptualization, Validation.\\
\textbf{Alberto Piatti}: Validation, Funding acquisition. \\
\textbf{Francesco Mondada}: Validation, Funding acquisition.\\
\textbf{Francesca Mangili}: Validation.\\
\textbf{Luca Maria Gambardella}: Funding acquisition.

\clearpage
\appendix

\section{Appendix}
\subsection{Research Instrument}

\begin{figure}[htb]
  \centering
  \begin{subfigure}[b]{0.33\textwidth}
    \centering
    \includegraphics[height=3.25cm]{pics/CTSKills_level1_Q1.PNG}
    \caption{Level 1, Q1}
  \end{subfigure}
  \hfill
  \begin{subfigure}[b]{0.33\textwidth}
    \centering
    \includegraphics[height=3.25cm]{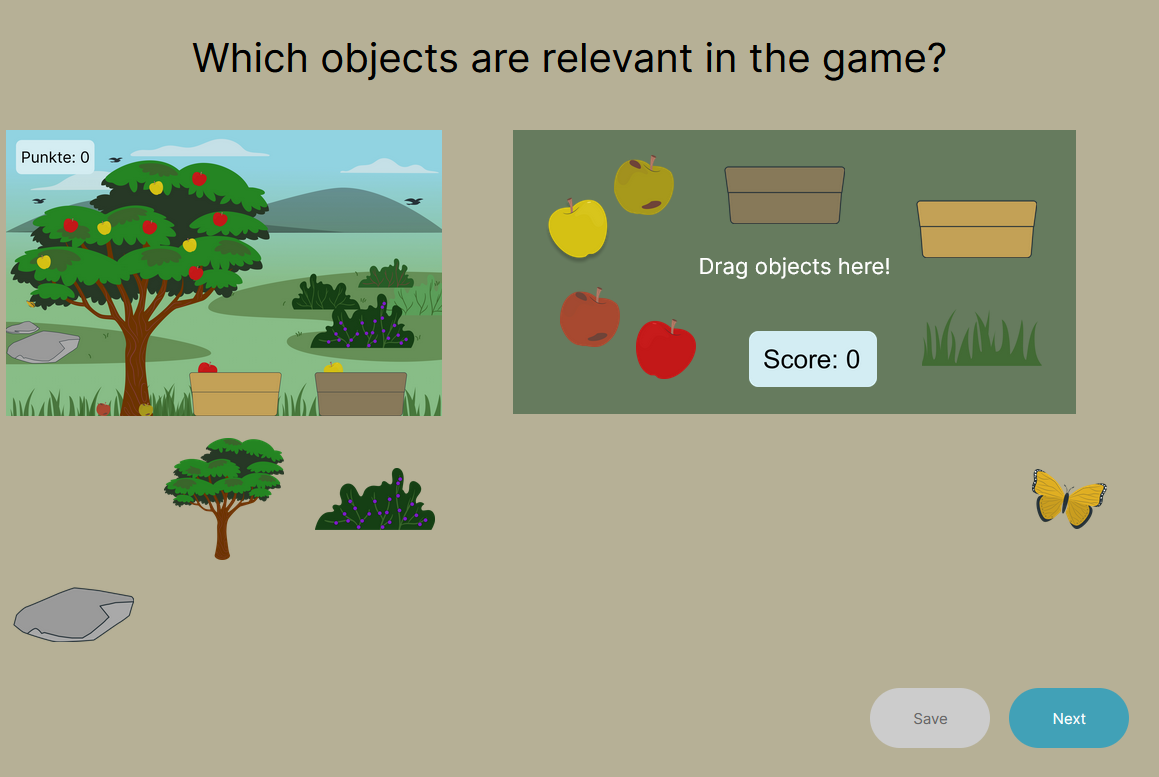}
    \caption{Level 2, Q1}
  \end{subfigure}
  \hfill
  \begin{subfigure}[b]{0.33\textwidth}
    \centering
    \includegraphics[height=3.25cm]{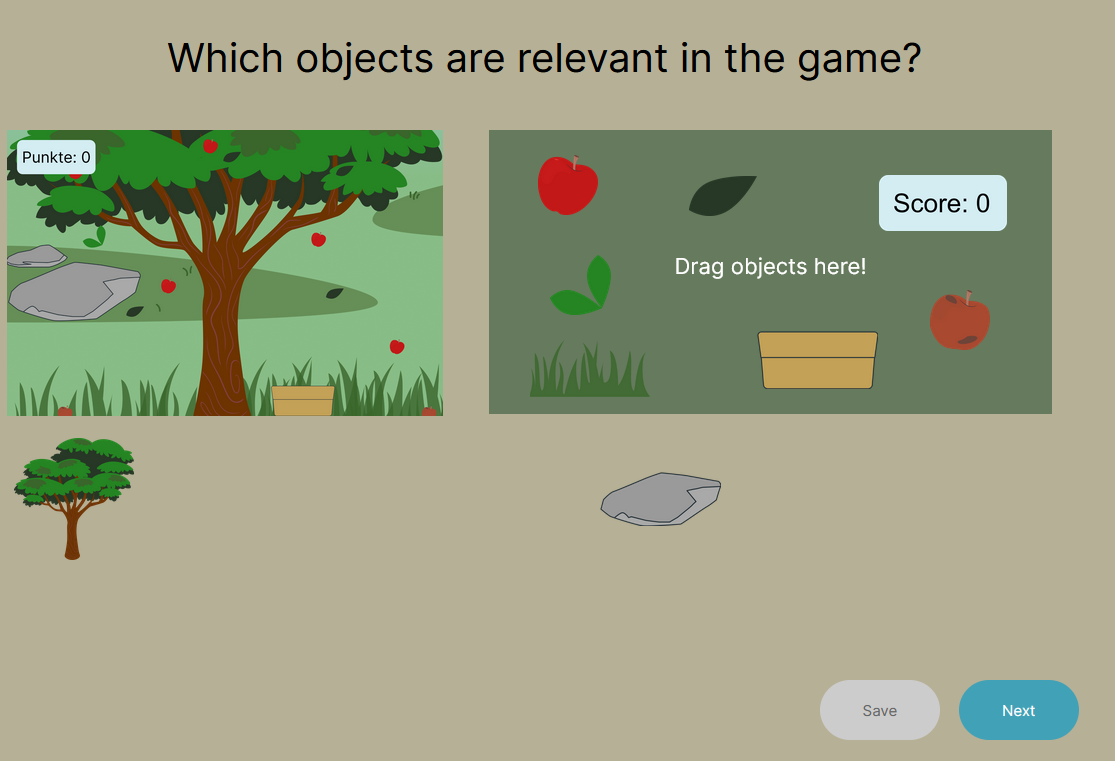}
    \caption{Level 3, Q1}
  \end{subfigure}
  \caption{Substantive decomposition and abstraction.}
  \Description{}
\end{figure}

\begin{figure}[htb]
  \centering
  \begin{subfigure}[b]{0.33\textwidth}
    \centering
    \includegraphics[height=3.25cm]{pics/CTSKills_level1_Q2.PNG}
    \caption{Level 1, Q2}
  \end{subfigure}
  \hfill
  \begin{subfigure}[b]{0.33\textwidth}
    \centering
    \includegraphics[height=3.25cm]{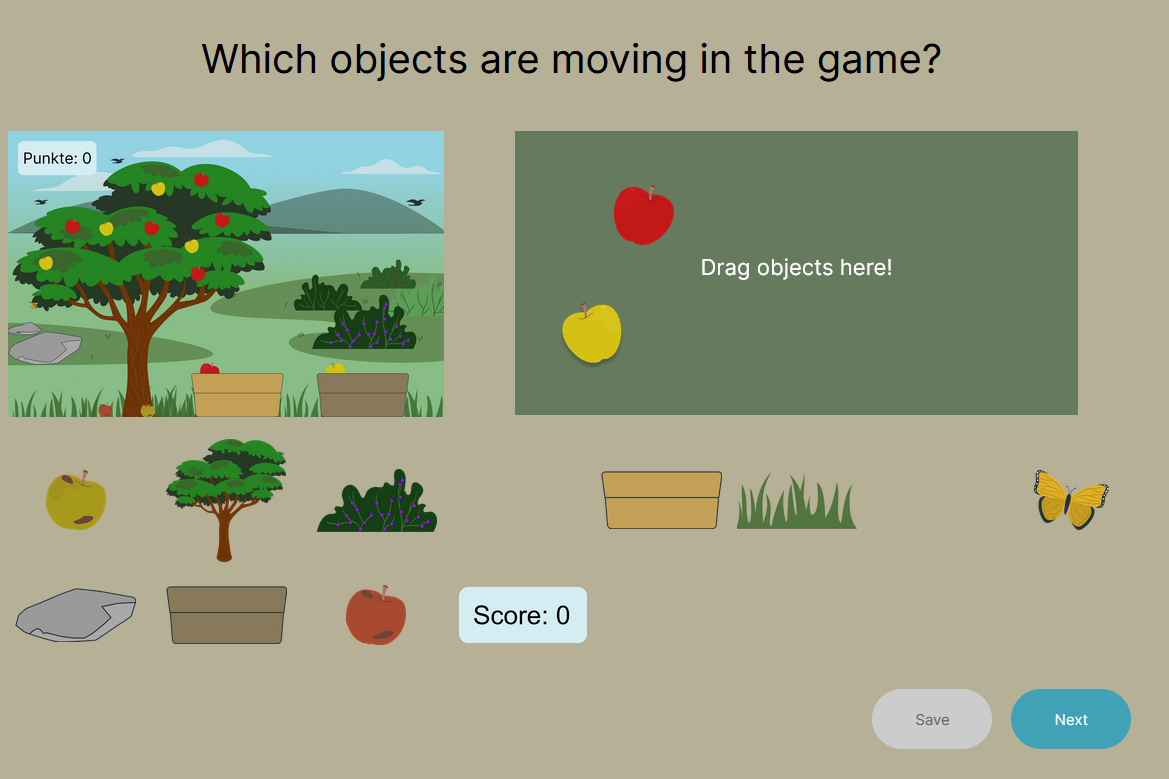}
    \caption{Level 2, Q2}
  \end{subfigure}
  \hfill
  \begin{subfigure}[b]{0.33\textwidth}
    \centering
    \includegraphics[height=3.25cm]{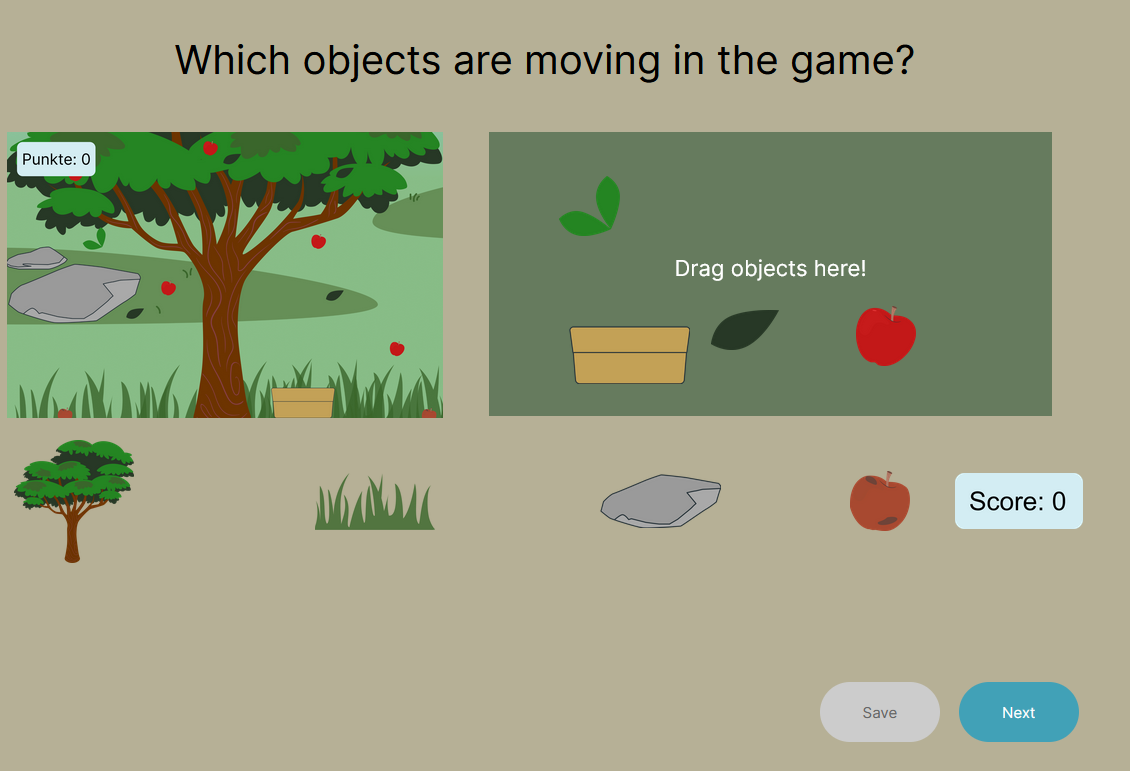}
    \caption{Level 3, Q2}
  \end{subfigure}
  \caption{Assigning properties to objects.}
  \Description{}
\end{figure}

\begin{figure}[htb]
  \centering
  \begin{subfigure}[b]{0.33\textwidth}
    \centering
    \includegraphics[height=3.25cm]{pics/CTSKills_level1_Q3.PNG}
    \caption{Level 1, Q3}
  \end{subfigure}
  \hfill
  \begin{subfigure}[b]{0.33\textwidth}
    \centering
    \includegraphics[height=3.25cm]{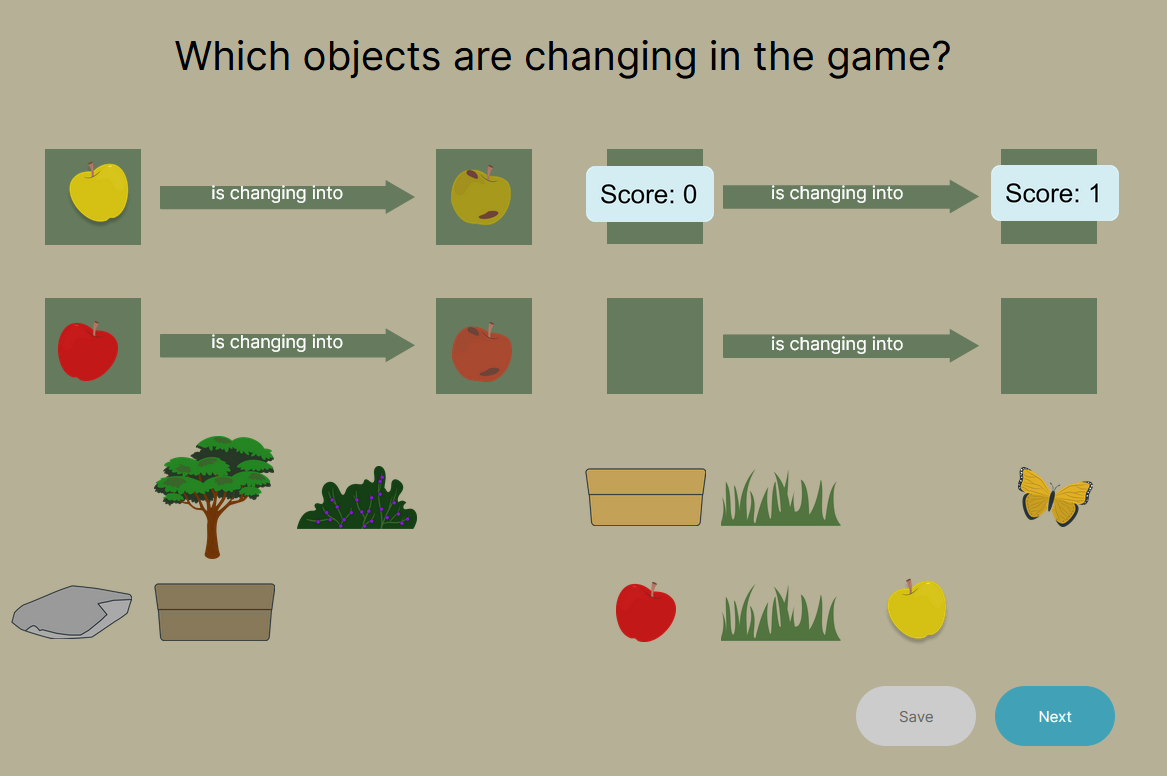}
    \caption{Level 2, Q3}
  \end{subfigure}
  \hfill
  \begin{subfigure}[b]{0.33\textwidth}
    \centering
    \includegraphics[height=3.25cm]{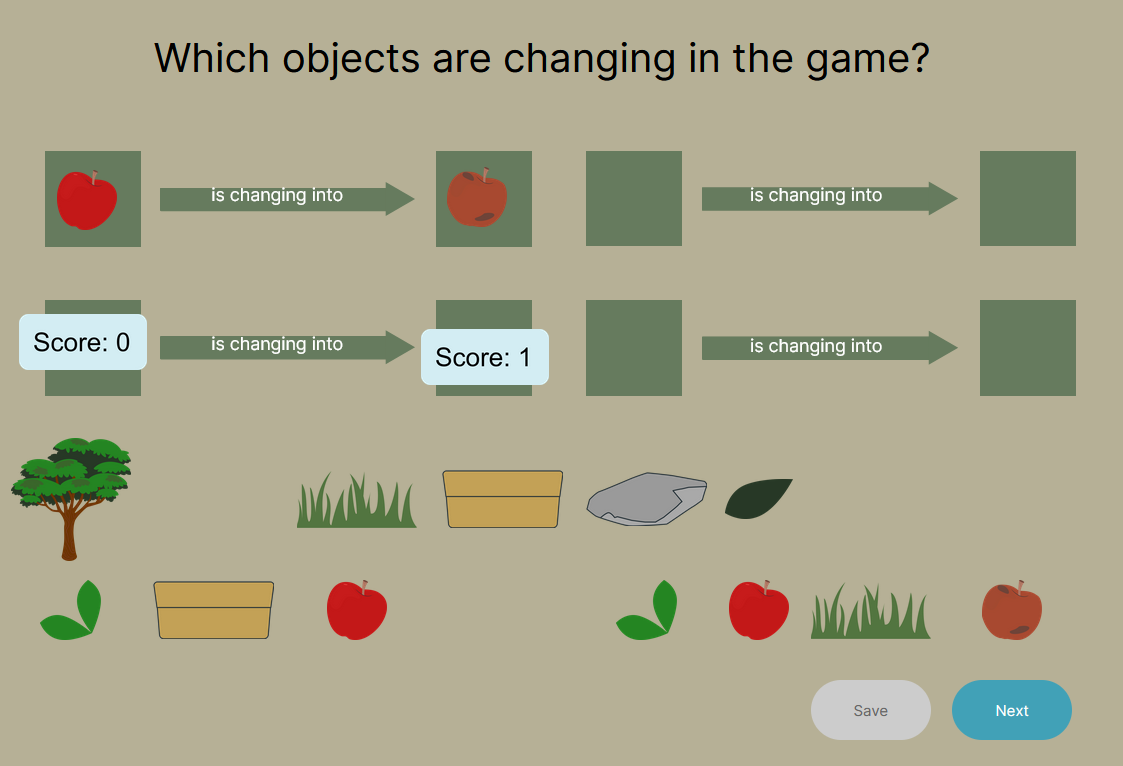}
    \caption{Level 3, Q3}
  \end{subfigure}
  \caption{Relational decomposition.}
  \Description{}
\end{figure}

\begin{figure}[htb]
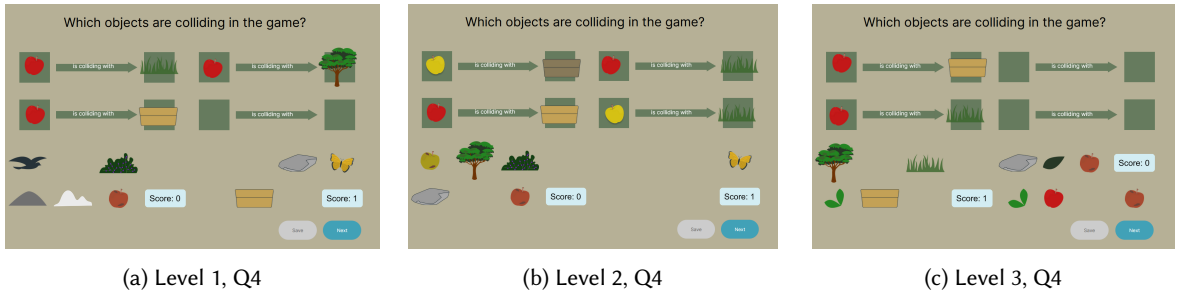

  \centering
  \begin{subfigure}[b]{0.33\textwidth}
    \centering
    \includegraphics[height=3.25cm]{pics/CTSKills_level1_Q4.PNG}
    \caption{Level 1, Q4}
    \label{fig:ctskills_level1_q4_appendix}
  \end{subfigure}
  \hfill
  \begin{subfigure}[b]{0.33\textwidth}
    \centering
    \includegraphics[height=3.25cm]{pics/CTSKills_level2_Q4.PNG}
    \caption{Level 2, Q4}
    \label{fig:ctskills_level2_q4_appendix}
  \end{subfigure}
  \hfill
  \begin{subfigure}[b]{0.33\textwidth}
    \centering
    \includegraphics[height=3.25cm]{pics/CTSKills_level3_Q4.PNG}
    \caption{Level 3, Q4}
    \label{fig:ctskills_level3_q4_appendix}
  \end{subfigure}
  \caption{Pattern recognition, abstraction and generalisation to make code reusable.}
  \label{fig:ctskills_questions4_appendix}
  \Description{}
\end{figure}

\subsection{Assessment}

Let $X$ represent the set of target items or pairs and $Y$ the set of non-target items or pairs. 
Table~\ref{tab:exercise_elements_appendix} shows the set of targets \(X\) and non-targets \(Y\) for each question and level.
It should also be noticed that, for questions Q3 and Q4, the number of non-target pairs \(|Y|\) is fixed at 4. This decision is due to the exercise interface allowing for the selection of up to 4 pairs on the screen, although fewer or more pairs could theoretically be chosen.

\begin{table}[ht]
\footnotesize
  \centering
  \caption{{Target and Non-Target Elements and Pairs.}
This table displays the types and quantities of target (\(X\)) and non-target (\(Y\)) elements across different questions (Q) and levels (L). For questions Q3 and Q4, only accepted target pairs are listed, while non-target pairs encompass a diverse range of combinations not explicitly listed. 
}
  \label{tab:exercise_elements_appendix}
  \begin{tabular}{cc|>{\arraybackslash}m{5.9cm}c|>{\arraybackslash}m{5.9cm}c}
    \toprule
    \textbf{Q} & \textbf{L} & \multicolumn{1}{>{\centering\arraybackslash}m{5.9cm}}{\textbf{Targets (\(X\))}} & \textbf{ \(|X|\)} & \multicolumn{1}{|>{\centering\arraybackslash}m{5.9cm}}{\textbf{Non-targets (\(Y\))}} & \textbf{\(|Y|\)}  \\
    \midrule
    1 & 1 & Apple red, Basket red, Score, Apple spoiled red, Grass & 5 & Rock, Bush, Bird, Butterfly, Mountain, Cloud & 6\\
    1 & 2 & Apple red, Basket red, Score, Apple spoiled red, Grass, Apple yellow, Basket yellow, Apple spoiled yellow & 8 &Rock, Bush, Butterfly & 3\\
    1 & 3 & Apple red, Basket red, Score, Apple spoiled red, Grass, Leaf light, Leaf dark & 7 & Rock &1 \\
    \hline
    2 & 1  & Apple red & 1& Basket red, Tree, Apple spoiled red, Grass, Rock, Bush, Bird, Butterfly, Mountain, Cloud & 10\\
    2 & 2 & Apple red, Apple yellow & 2 & Basket red, Tree, Apple spoiled red, Grass, Rock, Bush, Butterfly, Basket yellow, Apple spoiled yellow & 9\\
    2 & 3 & Apple red, Basket red, Leaf light, Leaf dark & 4 & Tree, Apple spoiled red, Grass, Rock & 4\\
    \hline
    3 & 1 & Apple red to Apple spoiled red, Score 0 to Score 1 & 2 & \textit{All other combinations} & 4 \\
    3 & 2 & Apple red to Apple spoiled red, Score 0 to Score 1, Apple yellow to Apple spoiled yellow & 3 & \textit{All other combinations} & 4 \\
    3 & 3 & Apple red to Apple spoiled red, Score 0 to Score 1 & 2 & \textit{All other combinations} & 4 \\
    \hline
    4 & 1 & Apple red and Basket red, Apple red and Grass, Apple spoiled red and Grass$^{*}$	& 2 & \textit{All other combinations} & 4 \\
    4 & 2 & Apple red and Basket red, Apple red and Grass, Apple yellow and Basket yellow, Apple yellow and Grass, Apple spoiled red and Grass$^{*}$, Apple spoiled yellow and Grass$^{*}$	& 4 & \textit{All other combinations} & 4 \\
    5 & 3 & Apple red and Basket red, Apple red and Grass, Apple spoiled red and Grass$^{*}$, Leaf dark and Grass$^{*}$, Leaf light and Grass$^{*}$& 2& \textit{All other combinations} & 4 \\
    \bottomrule
  \end{tabular}
  \\\vspace{0.5em}
 \begin{minipage}{.9\linewidth}
 $^{*}$ Optional pair, contributing a bonus of 0.5 to the score if selected.
 \end{minipage}
\end{table}

\FloatBarrier
\newpage
\mbox{~}
\newpage
\section{Results}

\begin{figure*}[!h]
  \centering
 \includegraphics[width=.85\linewidth]{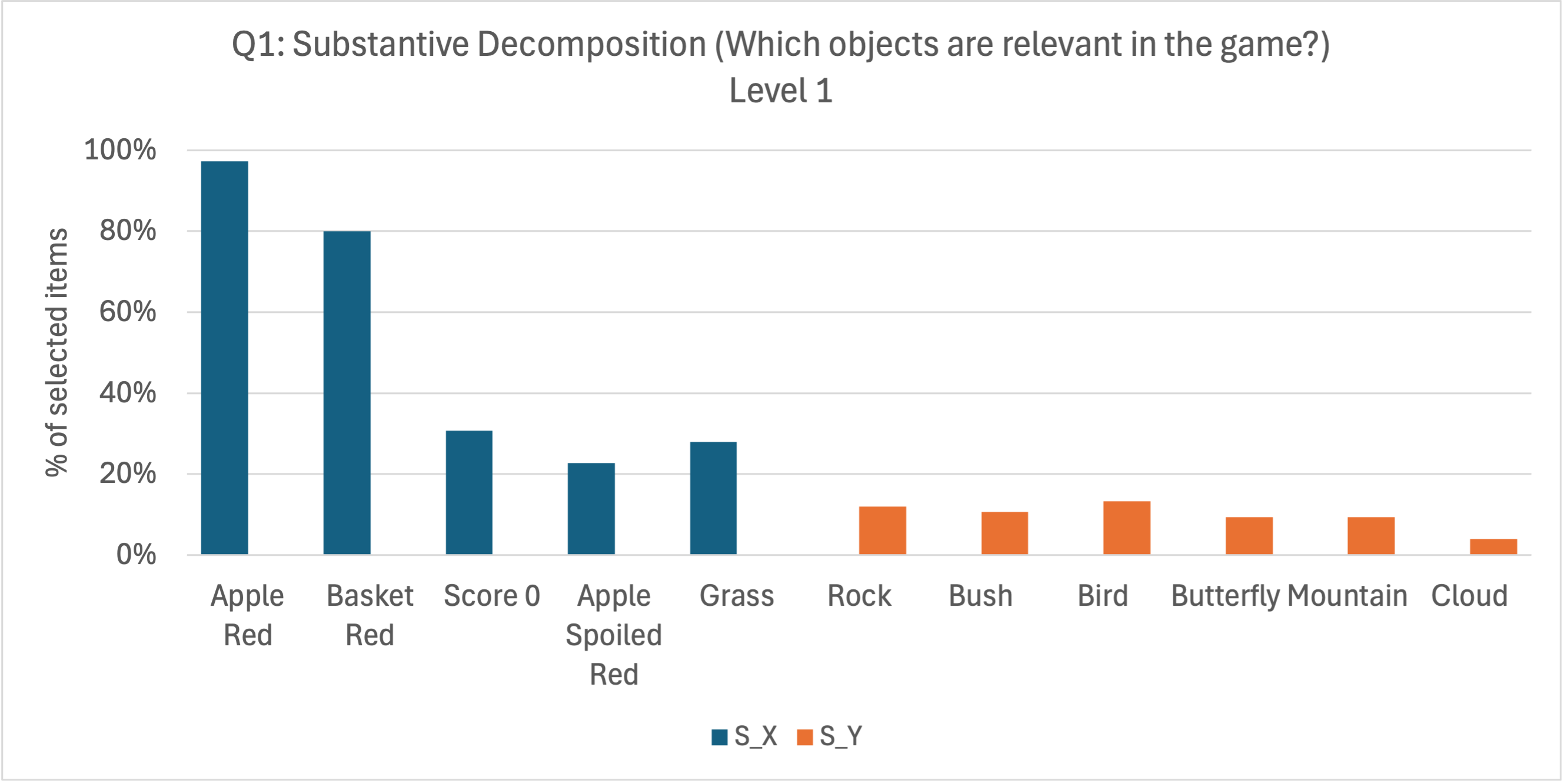}
  \Description{}
  \label{fig:ctskills_Q1_L1_results_appendix}
  \caption{Percentage of Times Each Item Was Selected as Relevant for Question 1 - Level 1.
  This chart illustrates the percentage of times each item was selected as relevant, differentiating between correctly selected targets $S_X$ (in blue) and incorrectly selected non-targets $S_Y$ (in orange).}
\end{figure*}
\begin{figure*}[!h]
  \centering
 \includegraphics[width=.85\linewidth]{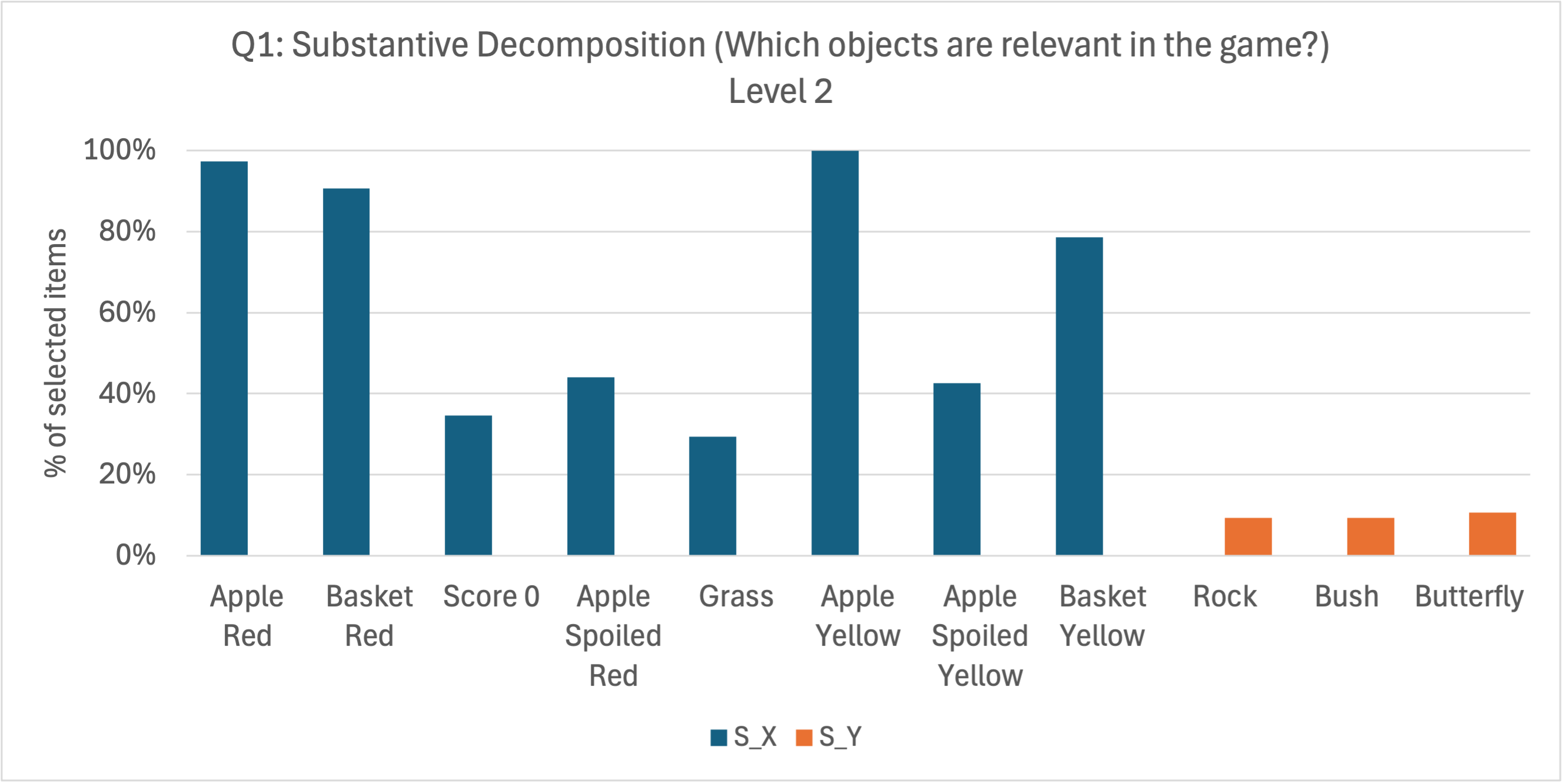}
  \Description{}
 \caption{Percentage of Times Each Item Was Selected as Relevant for Question 1 - Level 2.
  This chart illustrates the percentage of times each item was selected as relevant, differentiating between correctly selected targets $S_X$ (in blue) and incorrectly selected non-targets $S_Y$ (in orange).}
  \label{fig:ctskills_Q1_L2_results_appendix}
\end{figure*}

\begin{figure*}[!ht]
  \centering
 \includegraphics[width=.85\linewidth]{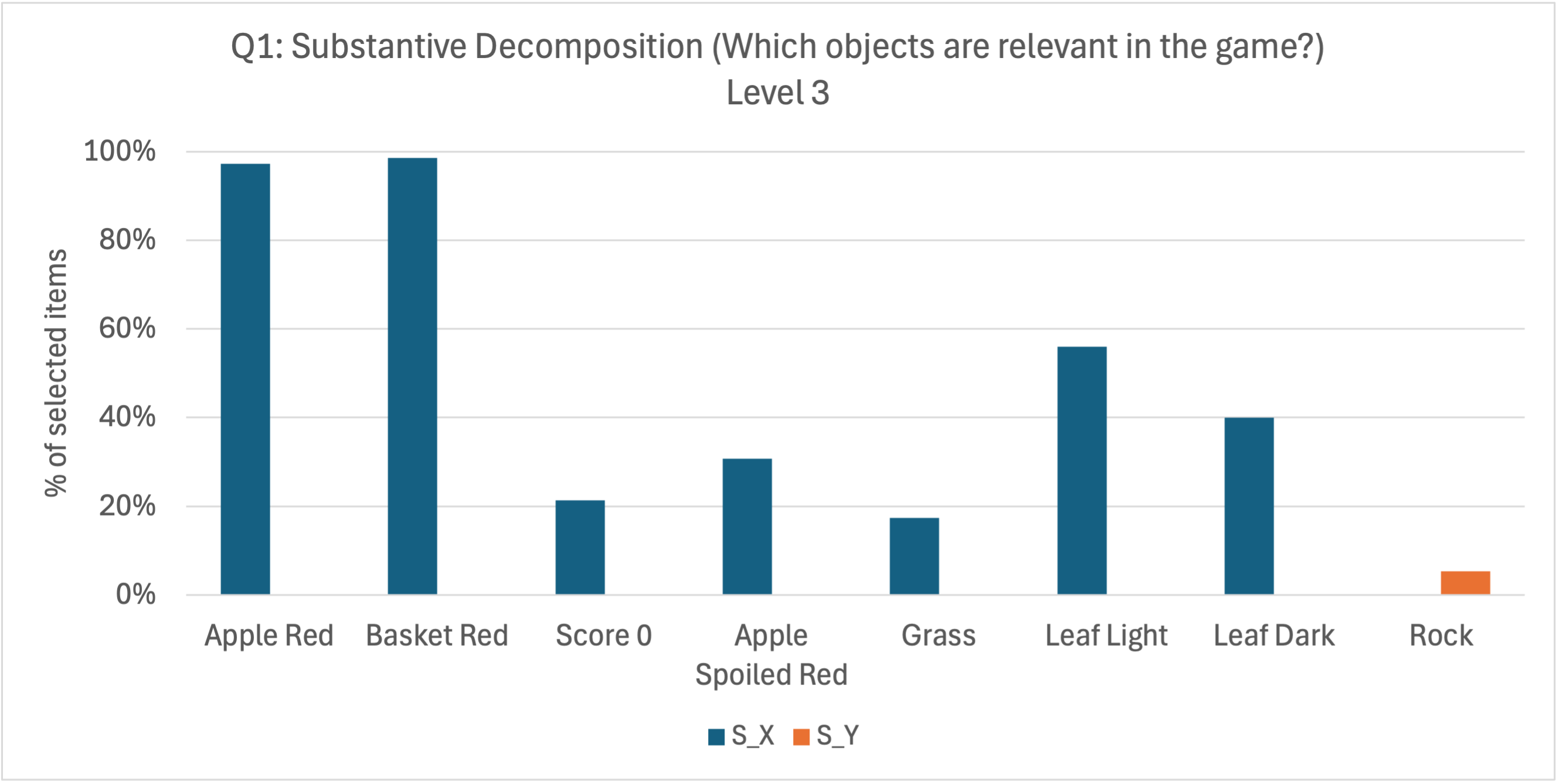}
  \Description{}
  \label{fig:ctskills_Q1_L3_results_appendix}
  \caption{Percentage of Times Each Item Was Selected as Relevant for Question 1 - Level 3.
  This chart illustrates the percentage of times each item was selected as relevant, differentiating between correctly selected targets $S_X$ (in blue) and incorrectly selected non-targets $S_Y$ (in orange).}
\end{figure*}

\begin{figure*}[!hb]
  \centering
 \begin{subfigure}[t]{.33\linewidth}
 \centering
 \includegraphics[height=4.7cm]{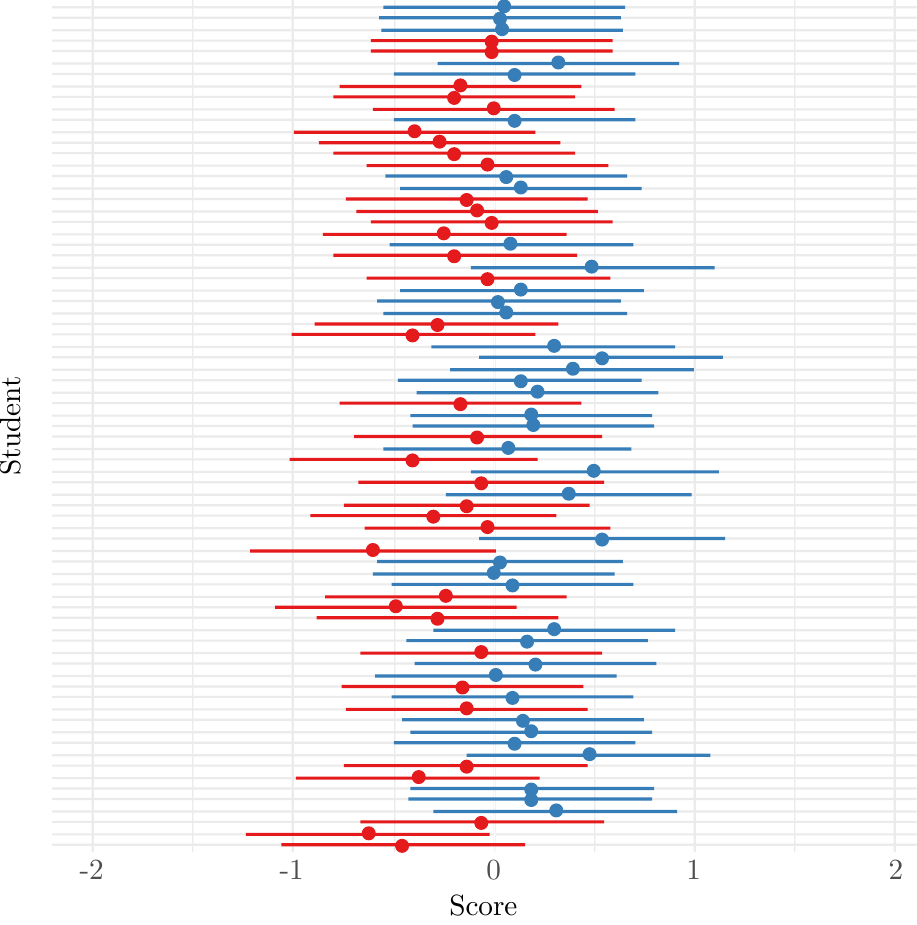}
 \caption{{Student Performance Variations}}
 \label{fig:re_student_appendix}
  \end{subfigure}
  \begin{subfigure}[t]{.33\linewidth}
 \centering
 \includegraphics[height=4.7cm]{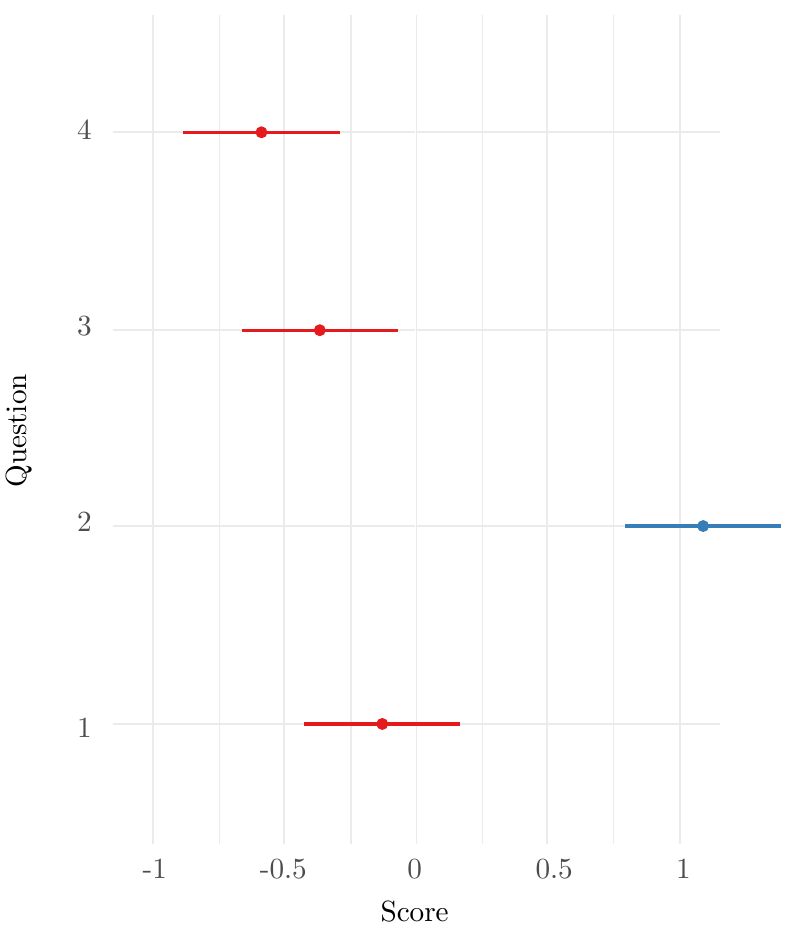}
 \caption{{Question Performance Variation}}
  \label{fig:re_question_appendix}
  \end{subfigure}
  \begin{subfigure}[t]{.33\linewidth}
 \centering
 \includegraphics[height=4.7cm]{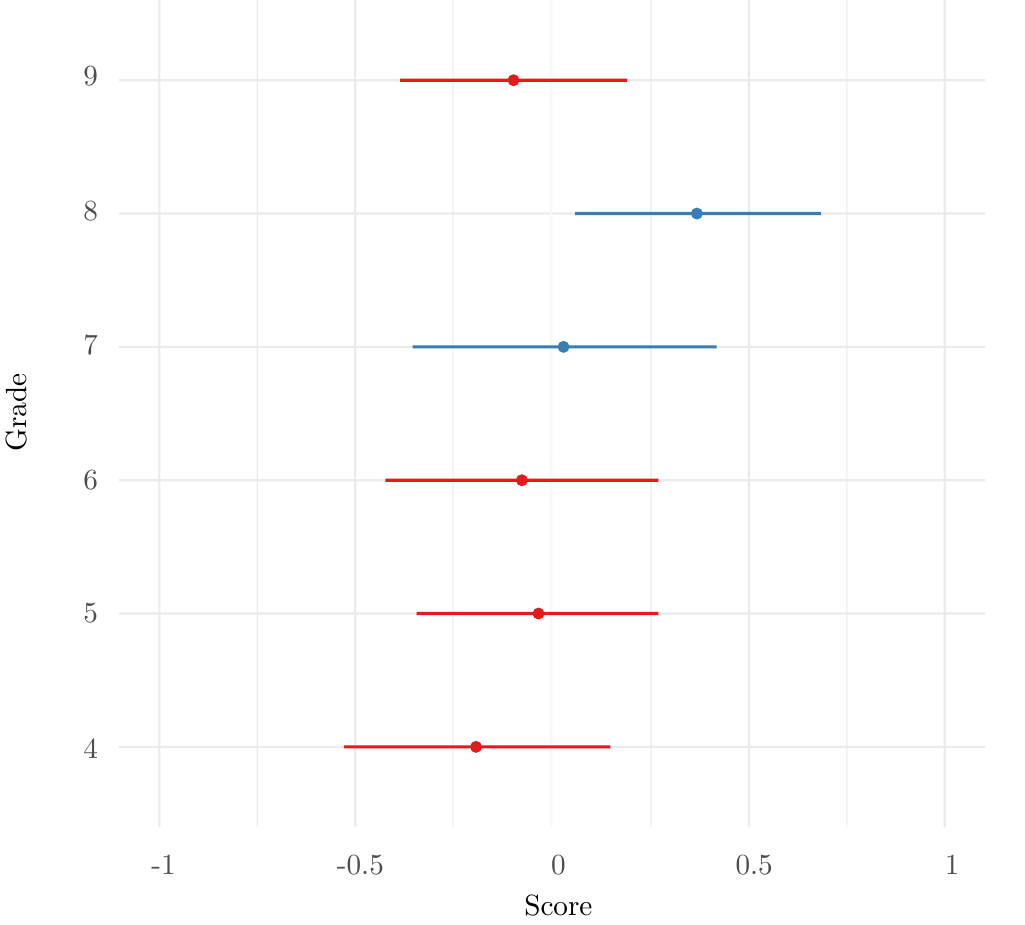}
 \caption{{Grade Performance Variation.}}
  \label{fig:re_grade_appendix}
  \end{subfigure}
  \caption{{Visualisation of the Random Effects.} 
  Each point represents the variable deviation from the average score, with blue indicating scores above average and red below. 
  Horizontal lines represent the estimates' confidence intervals.}
  \Description{}
  \label{fig:re_appendix}
\end{figure*}








\bibliographystyle{ACM-Reference-Format}
\bibliography{decomposition_paper}

\end{document}